\documentclass[12pt,preprint,sort]{aastex}
\usepackage{natbib}
\usepackage{lineno}
\linenumbers

\newcommand{\Fermi}{{\em Fermi}}
\newcommand{\Swift}{{\em Swift}}
\renewcommand{\L}{{\cal L}}

\renewcommand{\min}{{\rm min}}
\renewcommand{\max}{{\rm max}}
\newcommand{\Fmed}[1]{{\tilde{F}_{UL,{#1}}}}
\newcommand{\Epeak}{{E_{\rm pk}}}

\shorttitle{Constraining high-energy Emission with \Fermi}
\shortauthors{The \Fermi-LAT and \Fermi-GBM Collaborations}

\begin{document}

\title{Constraining the High-Energy Emission from Gamma-ray Bursts with \Fermi}

\author{
\begin{center}
The \Fermi\ Large Area Telescope Team
\end{center}
M.~Ackermann\altaffilmark{1}, 
M.~Ajello\altaffilmark{2}, 
L.~Baldini\altaffilmark{3}, 
G.~Barbiellini\altaffilmark{4,5}, 
M.~G.~Baring\altaffilmark{6}, 
K.~Bechtol\altaffilmark{2}, 
R.~Bellazzini\altaffilmark{3}, 
R.~D.~Blandford\altaffilmark{2}, 
E.~D.~Bloom\altaffilmark{2}, 
E.~Bonamente\altaffilmark{8,9}, 
A.~W.~Borgland\altaffilmark{2}, 
E.~Bottacini\altaffilmark{2}, 
A.~Bouvier\altaffilmark{10}, 
M.~Brigida\altaffilmark{11,12}, 
R.~Buehler\altaffilmark{2}, 
S.~Buson\altaffilmark{14,15}, 
G.~A.~Caliandro\altaffilmark{16}, 
R.~A.~Cameron\altaffilmark{2}, 
C.~Cecchi\altaffilmark{8,9}, 
E.~Charles\altaffilmark{2}, 
A.~Chekhtman\altaffilmark{17}, 
J.~Chiang\altaffilmark{2,18}, 
S.~Ciprini\altaffilmark{19,9}, 
R.~Claus\altaffilmark{2}, 
J.~Cohen-Tanugi\altaffilmark{20}, 
V.~Connaughton\altaffilmark{13,21}, 
S.~Cutini\altaffilmark{22}, 
F.~D'Ammando\altaffilmark{23,24}, 
F.~de~Palma\altaffilmark{11,12}, 
C.~D.~Dermer\altaffilmark{25}, 
E.~do~Couto~e~Silva\altaffilmark{2}, 
P.~S.~Drell\altaffilmark{2}, 
A.~Drlica-Wagner\altaffilmark{2}, 
C.~Favuzzi\altaffilmark{11,12}, 
Y.~Fukazawa\altaffilmark{27}, 
P.~Fusco\altaffilmark{11,12}, 
F.~Gargano\altaffilmark{12}, 
D.~Gasparrini\altaffilmark{22}, 
N.~Gehrels\altaffilmark{28}, 
S.~Germani\altaffilmark{8,9}, 
N.~Giglietto\altaffilmark{11,12}, 
F.~Giordano\altaffilmark{11,12}, 
M.~Giroletti\altaffilmark{29}, 
T.~Glanzman\altaffilmark{2}, 
J.~Granot\altaffilmark{30}, 
I.~A.~Grenier\altaffilmark{31}, 
J.~E.~Grove\altaffilmark{25}, 
S.~Guiriec\altaffilmark{13}, 
D.~Hadasch\altaffilmark{16}, 
Y.~Hanabata\altaffilmark{27}, 
A.~K.~Harding\altaffilmark{28}, 
E.~Hays\altaffilmark{28}, 
D.~Horan\altaffilmark{33}, 
G.~J\'ohannesson\altaffilmark{34}, 
J.~Kataoka\altaffilmark{35}, 
J.~Kn\"odlseder\altaffilmark{36,37}, 
D.~Kocevski\altaffilmark{2,38}, 
M.~Kuss\altaffilmark{3}, 
J.~Lande\altaffilmark{2}, 
F.~Longo\altaffilmark{4,5}, 
F.~Loparco\altaffilmark{11,12}, 
M.~N.~Lovellette\altaffilmark{25}, 
P.~Lubrano\altaffilmark{8,9}, 
M.~N.~Mazziotta\altaffilmark{12}, 
S.~McGlynn\altaffilmark{39}, 
P.~F.~Michelson\altaffilmark{2}, 
W.~Mitthumsiri\altaffilmark{2}, 
M.~E.~Monzani\altaffilmark{2}, 
E.~Moretti\altaffilmark{40,41,42}, 
A.~Morselli\altaffilmark{43}, 
I.~V.~Moskalenko\altaffilmark{2}, 
S.~Murgia\altaffilmark{2}, 
M.~Naumann-Godo\altaffilmark{31}, 
J.~P.~Norris\altaffilmark{44}, 
E.~Nuss\altaffilmark{20}, 
T.~Nymark\altaffilmark{40,41}, 
T.~Ohsugi\altaffilmark{45}, 
A.~Okumura\altaffilmark{2,46}, 
N.~Omodei\altaffilmark{2}, 
E.~Orlando\altaffilmark{2,32}, 
J.~H.~Panetta\altaffilmark{2}, 
D.~Parent\altaffilmark{47}, 
V.~Pelassa\altaffilmark{13}, 
M.~Pesce-Rollins\altaffilmark{3}, 
F.~Piron\altaffilmark{20}, 
G.~Pivato\altaffilmark{15}, 
J.~L.~Racusin\altaffilmark{28}, 
S.~Rain\`o\altaffilmark{11,12}, 
R.~Rando\altaffilmark{14,15}, 
S.~Razzaque\altaffilmark{47}, 
A.~Reimer\altaffilmark{7,2}, 
O.~Reimer\altaffilmark{7,2}, 
S.~Ritz\altaffilmark{10}, 
F.~Ryde\altaffilmark{40,41}, 
C.~Sgr\`o\altaffilmark{3}, 
E.~J.~Siskind\altaffilmark{48}, 
E.~Sonbas\altaffilmark{28,49,50}, 
G.~Spandre\altaffilmark{3}, 
P.~Spinelli\altaffilmark{11,12}, 
M.~Stamatikos\altaffilmark{28,51}, 
{\L}ukasz~Stawarz\altaffilmark{46,52}, 
D.~J.~Suson\altaffilmark{53}, 
H.~Takahashi\altaffilmark{45}, 
T.~Tanaka\altaffilmark{2}, 
J.~G.~Thayer\altaffilmark{2}, 
J.~B.~Thayer\altaffilmark{2}, 
L.~Tibaldo\altaffilmark{14,15}, 
M.~Tinivella\altaffilmark{3}, 
G.~Tosti\altaffilmark{8,9}, 
T.~Uehara\altaffilmark{27}, 
J.~Vandenbroucke\altaffilmark{2}, 
V.~Vasileiou\altaffilmark{20}, 
G.~Vianello\altaffilmark{2,54}, 
V.~Vitale\altaffilmark{43,55}, 
A.~P.~Waite\altaffilmark{2} \\
\begin{center}
The \Fermi\ Gamma-ray Burst Monitor Team 
\end{center}
V.~Connaughton\altaffilmark{13}, 
M.~S.~Briggs\altaffilmark{13,56}, 
S.~Guirec\altaffilmark{28}, 
A.~Goldstein\altaffilmark{13}, 
J.~M.~Burgess\altaffilmark{13}, 
P.~N.~Bhat\altaffilmark{13}, 
E.~Bissaldi\altaffilmark{7}, 
A. Camero-Arranz (50,58)
J.~Fishman\altaffilmark{13}, 
G.~Fitzpatrick\altaffilmark{26}, 
S.~Foley\altaffilmark{26,32}, 
D.~Gruber\altaffilmark{32}, 
P. Jenke\altaffilmark{58}, 
R.~M.~Kippen\altaffilmark{57}, 
C.~Kouveliotou\altaffilmark{58}, 
S.~McBreen\altaffilmark{26,32}, 
C.~Meegan\altaffilmark{50}, 
W.~S.~Paciesas\altaffilmark{13}, 
R.~Preece\altaffilmark{13}, 
A.~Rau\altaffilmark{32}, 
D.~Tierney\altaffilmark{26}, 
A.~J.~van~der~Horst\altaffilmark{58,59}, 
A.~von~Kienlin\altaffilmark{32}, 
C.~Wilson-Hodge\altaffilmark{58},
S.~Xiong\altaffilmark{13}
}
\altaffiltext{1}{Deutsches Elektronen Synchrotron DESY, D-15738 Zeuthen, Germany}
\altaffiltext{2}{W. W. Hansen Experimental Physics Laboratory, Kavli Institute for Particle Astrophysics and Cosmology, Department of Physics and SLAC National Accelerator Laboratory, Stanford University, Stanford, CA 94305, USA}
\altaffiltext{3}{Istituto Nazionale di Fisica Nucleare, Sezione di Pisa, I-56127 Pisa, Italy}
\altaffiltext{4}{Istituto Nazionale di Fisica Nucleare, Sezione di Trieste, I-34127 Trieste, Italy}
\altaffiltext{5}{Dipartimento di Fisica, Universit\`a di Trieste, I-34127 Trieste, Italy}
\altaffiltext{6}{Rice University, Department of Physics and Astronomy, MS-108, P. O. Box 1892, Houston, TX 77251, USA}
\altaffiltext{7}{Institut f\"ur Astro- und Teilchenphysik and Institut f\"ur Theoretische Physik, Leopold-Franzens-Universit\"at Innsbruck, A-6020 Innsbruck, Austria}
\altaffiltext{8}{Istituto Nazionale di Fisica Nucleare, Sezione di Perugia, I-06123 Perugia, Italy}
\altaffiltext{9}{Dipartimento di Fisica, Universit\`a degli Studi di Perugia, I-06123 Perugia, Italy}
\altaffiltext{10}{Santa Cruz Institute for Particle Physics, Department of Physics and Department of Astronomy and Astrophysics, University of California at Santa Cruz, Santa Cruz, CA 95064, USA}
\altaffiltext{11}{Dipartimento di Fisica ``M. Merlin" dell'Universit\`a e del Politecnico di Bari, I-70126 Bari, Italy}
\altaffiltext{12}{Istituto Nazionale di Fisica Nucleare, Sezione di Bari, 70126 Bari, Italy}
\altaffiltext{13}{Center for Space Plasma and Aeronomic Research (CSPAR), University of Alabama in Huntsville, Huntsville, AL 35899, USA}
\altaffiltext{14}{Istituto Nazionale di Fisica Nucleare, Sezione di Padova, I-35131 Padova, Italy}
\altaffiltext{15}{Dipartimento di Fisica ``G. Galilei", Universit\`a di Padova, I-35131 Padova, Italy}
\altaffiltext{16}{Institut de Ci\`encies de l'Espai (IEEE-CSIC), Campus UAB, 08193 Barcelona, Spain}
\altaffiltext{17}{Artep Inc., 2922 Excelsior Springs Court, Ellicott City, MD 21042, resident at Naval Research Laboratory, Washington, DC 20375, USA}
\altaffiltext{18}{email: jchiang@slac.stanford.edu}
\altaffiltext{19}{ASI Science Data Center, I-00044 Frascati (Roma), Italy}
\altaffiltext{20}{Laboratoire Univers et Particules de Montpellier, Universit\'e Montpellier 2, CNRS/IN2P3, Montpellier, France}
\altaffiltext{21}{email: connauv@uah.edu}
\altaffiltext{22}{Agenzia Spaziale Italiana (ASI) Science Data Center, I-00044 Frascati (Roma), Italy}
\altaffiltext{23}{IASF Palermo, 90146 Palermo, Italy}
\altaffiltext{24}{INAF-Istituto di Astrofisica Spaziale e Fisica Cosmica, I-00133 Roma, Italy}
\altaffiltext{25}{Space Science Division, Naval Research Laboratory, Washington, DC 20375-5352, USA}
\altaffiltext{26}{University College Dublin, Belfield, Dublin 4, Ireland}
\altaffiltext{27}{Department of Physical Sciences, Hiroshima University, Higashi-Hiroshima, Hiroshima 739-8526, Japan}
\altaffiltext{28}{NASA Goddard Space Flight Center, Greenbelt, MD 20771, USA}
\altaffiltext{29}{INAF Istituto di Radioastronomia, 40129 Bologna, Italy}
\altaffiltext{30}{Centre for Astrophysics Research, Science and Technology Research Institute, University of Hertfordshire, Hatfield AL10 9AB, UK}
\altaffiltext{31}{Laboratoire AIM, CEA-IRFU/CNRS/Universit\'e Paris Diderot, Service d'Astrophysique, CEA Saclay, 91191 Gif sur Yvette, France}
\altaffiltext{32}{Max-Planck Institut f\"ur extraterrestrische Physik, 85748 Garching, Germany}
\altaffiltext{33}{Laboratoire Leprince-Ringuet, \'Ecole polytechnique, CNRS/IN2P3, Palaiseau, France}
\altaffiltext{34}{Science Institute, University of Iceland, IS-107 Reykjavik, Iceland}
\altaffiltext{35}{Research Institute for Science and Engineering, Waseda University, 3-4-1, Okubo, Shinjuku, Tokyo 169-8555, Japan}
\altaffiltext{36}{CNRS, IRAP, F-31028 Toulouse cedex 4, France}
\altaffiltext{37}{GAHEC, Universit\'e de Toulouse, UPS-OMP, IRAP, Toulouse, France}
\altaffiltext{38}{email: kocevski@stanford.edu}
\altaffiltext{39}{Exzellenzcluster Universe, Technische Universit\"at M\"unchen, D-85748 Garching, Germany}
\altaffiltext{40}{Department of Physics, Royal Institute of Technology (KTH), AlbaNova, SE-106 91 Stockholm, Sweden}
\altaffiltext{41}{The Oskar Klein Centre for Cosmoparticle Physics, AlbaNova, SE-106 91 Stockholm, Sweden}
\altaffiltext{42}{email: moretti@particle.kth.se}
\altaffiltext{43}{Istituto Nazionale di Fisica Nucleare, Sezione di Roma ``Tor Vergata", I-00133 Roma, Italy}
\altaffiltext{44}{Department of Physics, Boise State University, Boise, ID 83725, USA}
\altaffiltext{45}{Hiroshima Astrophysical Science Center, Hiroshima University, Higashi-Hiroshima, Hiroshima 739-8526, Japan}
\altaffiltext{46}{Institute of Space and Astronautical Science, JAXA, 3-1-1 Yoshinodai, Chuo-ku, Sagamihara, Kanagawa 252-5210, Japan}
\altaffiltext{47}{Center for Earth Observing and Space Research, College of Science, George Mason University, Fairfax, VA 22030, resident at Naval Research Laboratory, Washington, DC 20375, USA}
\altaffiltext{48}{NYCB Real-Time Computing Inc., Lattingtown, NY 11560-1025, USA}
\altaffiltext{49}{Ad{\i}yaman University, 02040 Ad{\i}yaman, Turkey}
\altaffiltext{50}{Universities Space Research Association (USRA), Columbia, MD 21044, USA}
\altaffiltext{51}{Department of Physics, Center for Cosmology and Astro-Particle Physics, The Ohio State University, Columbus, OH 43210, USA}
\altaffiltext{52}{Astronomical Observatory, Jagiellonian University, 30-244 Krak\'ow, Poland}
\altaffiltext{53}{Department of Chemistry and Physics, Purdue University Calumet, Hammond, IN 46323-2094, USA}
\altaffiltext{54}{Consorzio Interuniversitario per la Fisica Spaziale (CIFS), I-10133 Torino, Italy}
\altaffiltext{55}{Dipartimento di Fisica, Universit\`a di Roma ``Tor Vergata", I-00133 Roma, Italy}
\altaffiltext{56}{email: michael.briggs@nasa.gov}
\altaffiltext{57}{Los Alamos National Laboratory, Los Alamos, NM 87545, USA}
\altaffiltext{58}{NASA Marshall Space Flight Center, Huntsville, AL 35812, USA}
\altaffiltext{59}{NASA Postdoctoral Program Fellow, USA}
\altaffiltext{60}{email: valerie@nasa.gov}

\begin{abstract}

We examine 288 GRBs detected by the \Fermi\ Gamma-ray Space Telescope's Gamma-ray Burst Monitor (GBM) that fell within the field-of-view of {\em Fermi's} Large Area Telescope (LAT) during the first 2.5 years of observations, which showed no evidence for emission
above 100 MeV.  We report the photon flux upper limits in the 0.1$-$10
GeV range during the prompt emission phase as well as for fixed 30 s and 100
s integrations starting from the trigger time for each burst.  We
compare these limits with the fluxes that would be expected from
extrapolations of spectral fits presented in the first GBM spectral
catalog and infer that roughly half of the GBM-detected bursts either
require spectral breaks between the GBM and LAT energy bands or have
intrinsically steeper spectra above the peak of the $\nu F_{\nu}$
spectra ($E_{\rm pk}$).  In order to distinguish between these two
scenarios, we perform joint GBM and LAT spectral fits to the 30
brightest GBM-detected bursts and find that a majority of these bursts are
indeed softer above $E_{\rm pk}$ than would be inferred from fitting
the GBM data alone.  Approximately 20\% of this spectroscopic
subsample show statistically significant evidence for a cut-off in
their high-energy spectra, which if assumed to be due to
$\gamma\gamma$ attenuation, places limits on the maximum Lorentz
factor associated with the relativistic outflow producing this
emission.  All of these latter bursts have maximum Lorentz factor
estimates that are well below the minimum Lorentz factors calculated
for LAT-detected GRBs, revealing a wide distribution in the bulk
Lorentz factor of GRB outflows and indicating that LAT-detected bursts
may represent the high end of this distribution.

\end{abstract}


\section{Introduction}
 
Observations by the \Fermi\ Gamma-ray Space Telescope have dramatically increased our knowledge of the broad-band spectra of gamma-ray bursts (GRBs). The Gamma-ray Burst Monitor (GBM) on board \Fermi\ has detected over 700 GRBs in roughly 3 years of triggered operations.  Of these bursts, 29 have been detected at energies $>$\,100\,MeV by {\em Fermi's} Large Area Telescope (LAT); and five of these bursts: GRB~080916C, GRB~090510, GRB~090328, GRB~090902B, and GRB~090926A, have been detected at energies $>\,10$\,GeV.  The high-energy emission from the majority of these bursts show evidence for being consistent with the high-energy component of the smoothly joined broken power-law, commonly referred to as the Band spectrum \citep{Band93}, that has been observed in the GBM energy range.  Three of these bursts: GRB~090510 \citep{Ackermann10}, GRB~090902B \citep{Abdo09}, and GRB~090926A \citep{Ackermann11}, though, exhibit an additional hard spectral component that is distinct from the continuum emission observed at sub-MeV energies.   
 
Similar high-energy emission above 100 MeV was detected by the Energetic Gamma-Ray Experiment Telescope (\emph{EGRET}) onboard the Compton Gamma-Ray Observatory and by the \emph{AGILE} spacecraft \citep{DelMonte11}.  The prompt high-energy emission detected by \emph{EGRET} from GRB~930131 \citep{Sommer94,Kouveliotou94} and GRB~940217 \citep{Hurley94}, was consistent with an extrapolation of the GRB spectrum as measured by the Burst And Transient Source Experiment (BATSE) in the 25\,keV$-$2\,MeV energy range.  \emph{EGRET} observations of GRB~941017 \citep{Gonzalez03}, on the other hand, showed evidence for an additional hard spectral component that extended up to 200\,MeV, the first such detection in a GRB spectrum.

Unlike these previous detections by \emph{EGRET}, many of the LAT detected bursts have measured redshifts, made possible through X-ray
localizations by the \Swift\ spacecraft \citep{Gehrels04} and ground-based follow-up observations of their long-lived afterglow emission. The high-energy detections, combined with the redshift to these GRBs, have shed new light into the underlying physics of this emission.  At a redshift of $z=0.903$ \citep{McBreen10}, the detection of GeV photons from GRB~090510 indicates a minimum bulk Lorentz factor of $\Gamma_{\gamma\gamma,\min} \sim 1200$ in order for the observed gamma rays to have avoided attenuation due to electron-positron pair production \citep{GRB090510_physics}.  Furthermore, a spectral cut-off at $\sim 1.4$ GeV is quite evident in the high-energy component of GRB~090926A, which if interpreted as opacity due to $\gamma\gamma$ attenuation within the emitting region, allows for a direct estimate of the bulk Lorentz factor of $\Gamma \sim 200-$700 for the outflow producing the emission \citep{090926A}.

Perhaps equally important for unraveling the nature of the prompt
emission is the lack of a significant detection above 100 MeV for the
majority of the GRBs detected by the GBM.  The LAT instrument has
detected roughly 8$\%$ of the GBM-triggered GRBs that have occurred
within the LAT field-of-view (FOV).  This detection rate places limits on the
ubiquity of the extra high-energy components detected by LAT, \emph{EGRET}, and \emph{AGILE}.  Such a component would be a natural consequence of synchrotron emission from relativistic electrons in an internal shock scenario, but, for example, might be suppressed in Poynting flux dominated models
(e$.$g$.$, see \citet{Fan08}). Therefore, a systematic analysis of the non-detections of high-energy components in GBM-detected GRBs may significantly help to discriminate between various prompt emission mechanisms.  Furthermore, the lack of a detection by the LAT of GBM-detected GRBs with particularly hard spectra points to intrinsic spectral cut-offs and/or curvature at high energies, giving us further insight into the physical properties of the emitting region.
 
In this paper, we examine the GBM-detected bursts that fell within the LAT field-of-view at the time of trigger during the first 2.5 years of observations which showed no evidence for emission above 100 MeV.  We report the photon flux upper limits in the 0.1$-$10\,GeV band during the prompt emission phase and for 30\,s and 100\,s integrations starting from the trigger time for each burst.  We then compare these upper limits with the fluxes that would be expected from extrapolations of spectral fits presented in the first GBM spectral catalog (Goldstein et al$.$, in press) in order to determine how well measurements of the $\la$\,MeV properties of GRBs can predict detections at $>100$\,MeV energies. 

We find that roughly half of the GBM detected bursts either require spectral breaks or have intrinsically steeper spectra in order to explain their non-detections by the LAT.  We distinguish between these two scenarios by performing joint GBM and LAT spectral fits to a subset of the 30 brightest bursts, as seen by the GBM that were simultaneously in the LAT field of view.  We find that while a majority of these bursts have spectra that are softer above the peak of the $\nu F_{\nu}$ spectra ($E_{\rm pk}$) than would be inferred from fitting the GBM data alone, a subset of bright bursts have a svstatistically significant high-energy spectral cut-off similar to the spectral break reported for GRB~090926A \citep{090926A}.  Finally, we use our joint GBM and LAT spectral fits in conjunction with the LAT non-detections at 100 MeV to place limits on the maximum Lorentz factor for these GRBs which show evidence for intrinsic spectral breaks 

The paper is structured as follows: In section~\ref{sec:InstrumentOverview}, we review the characteristics of the GBM and LAT instruments, and in
section~\ref{sec:SampleDefinition}, we define the GRB samples
considered in this work.  In section~\ref{sec:Analysis}, we describe
the analysis we perform to quantify the significance of the LAT
non-detections; we present the results in section~\ref{sec:Results},
and discuss the implications they have on our understanding of the
properties associated with the prompt gamma-ray emission in
section~\ref{sec:Discussion}.
 
\section{The LAT and GBM Instruments} \label{sec:InstrumentOverview}

The \Fermi\ Gamma-ray Space Telescope carries the Gamma-ray Burst
Monitor \citep{Meegan:09} and the Large Area Telescope \citep{Atwood:09}. The GBM has 14
scintillation detectors that together view the entire unocculted
sky. Triggering and localization are performed using 12 sodium iodide
(NaI) and 2 bismuth germanate (BGO) detectors with different orientations placed around the
spacecraft. The two BGO scintillators are placed on
opposite sides of the spacecraft so that at least one detector is in view for any direction on the sky.  GBM spectroscopy uses both the NaI and BGO
detectors, sensitive between 8 keV and 1 MeV, and 150 keV and 40 MeV,
respectively, so that their combination provides an unprecedented 4
decades of energy coverage with which to perform spectroscopic studies of
GRBs. 

The LAT is a pair conversion telescope comprising a $4\times4$ array of silicon strip trackers and cesium
iodide (CsI) calorimeters covered by a segmented anti-coincidence
detector (ACD) to reject charged-particle background events. The LAT
covers the energy range from 20\,MeV to more than 300\,GeV with a
field-of-view of $\sim 2.4$ steradians. The dead time per event of the LAT is
nominally 26.50\,$\mu$s for most events, although about 10$\%$ of the event read outs include more calibration data, which engender longer dead times.  This dead time is 4 orders of magnitude shorter than that of \emph{EGRET}. This is crucial for observations of high-intensity
transient events such as GRBs.  The LAT triggers on many more
background events than celestial gamma rays. Onboard background
rejection is supplemented on the ground using event class selections
that accommodate the broad range of sources of interest.

\newcommand{\Tzero}{\mbox{$T_0$}}

\section{Sample Definition} \label{sec:SampleDefinition}

\begin{figure*}
\includegraphics[width=7in,angle=0]{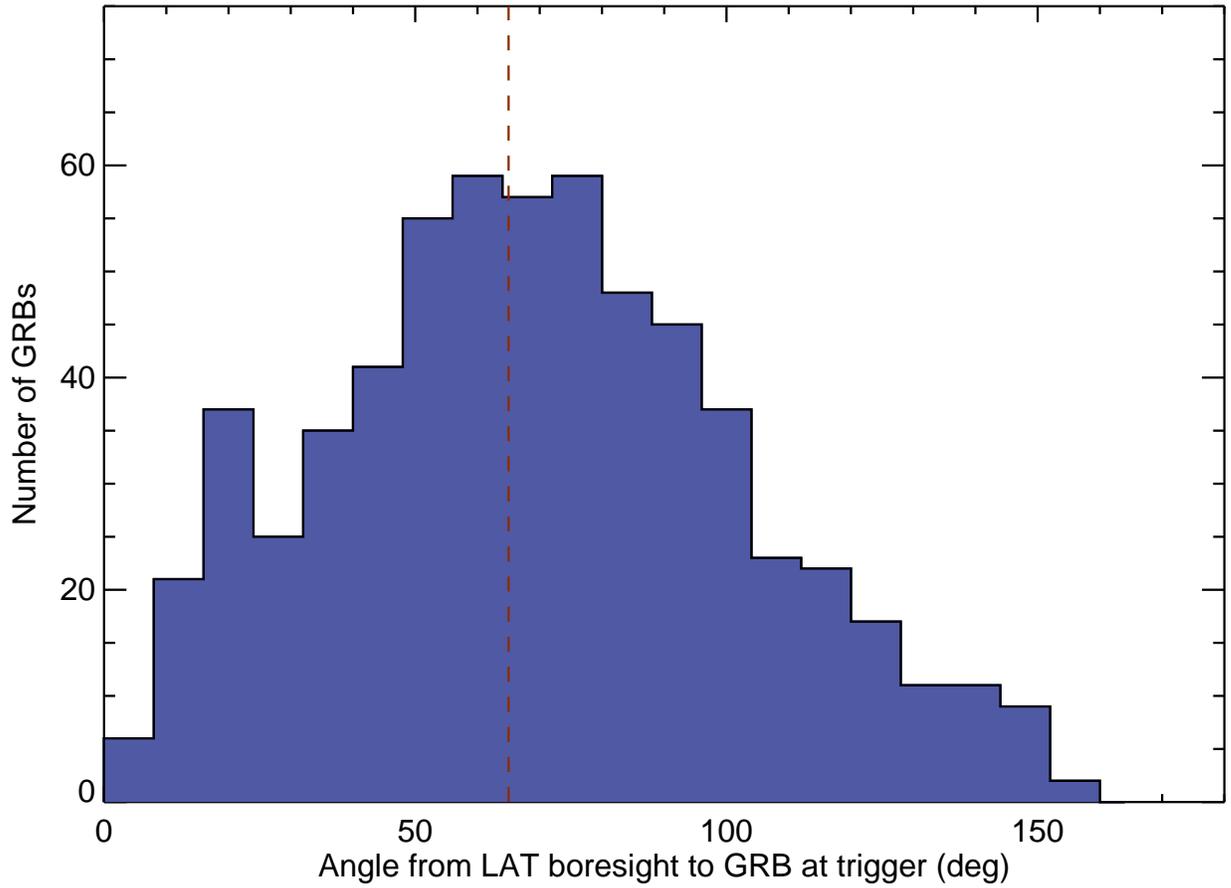}
\caption{The distribution of LAT off-axis angles of the 620 bursts
  that triggered the GBM from 2008 August 4th to 2011 January 1st.  The red
  dashed line at an off-axis angle of $65^\circ$ indicates the
  nominal boundary of the LAT FOV.  A total of 288 bursts (46\% of all detected bursts) fell within the LAT FOV over this period.}
\label{boresightangledist}
\end{figure*}

We compiled a sample of all GRBs detected by the GBM between the beginning of normal science operations  of the \Fermi\ mission
on 2008 August 4th up to 2011 January 1st, yielding a total of 620 GRBs.  Of
these, 288 bursts fell within 65$^\circ$ of the LAT z-axis (or
boresight) at the time of GBM trigger, which we define as the LAT FOV.  Bursts
detected at angles greater than $65^\circ$ at the time of the GBM trigger were not considered for
this analysis, due to the greatly reduced sensitivity of the
instrument for such large off-axis angles.  A plot of the distribution
of the LAT boresight angles at trigger time, \Tzero, for all 620
bursts is shown in Figure~\ref{boresightangledist}.  Roughly half
(46$\%$) of the GBM-detected GRBs fell within the LAT FOV at \Tzero,
as expected given the relative sky coverage of the two instruments.  These bursts make up the sample for which the photon flux upper limits described in the next section have been calculated. A complete list of the 288 bursts in the sample, their positions, their
durations, and their LAT boresight angles is given in
Table~\ref{Table:SampleDefinition}. 

We defined a subsample of 92 bursts which had a rate trigger greater
than 75 counts\, s$^{-1}$ in at least one of the two BGO detectors.  This criteria is similar to the one adopted by \citet{Bissaldi11} in their analysis of the brightest GBM detected bursts in the first year of observations. Hereafter, we refer to these 92 bursts as the ``bright BGO subsample''; it
comprises likely candidates for which it would be possible to find
evidence of spectral curvature above the upper boundary of the nominal BGO energy window of $\sim 40$\,MeV.
Finally, we define our so-called ``spectroscopic subsample'' as
the 30 bursts (of the bright BGO subsample) that have sufficient
counts at higher energies to allow for the $\beta$ index of a Band
function fit to be determined with standard errors $\le 0.5$.  This
spectroscopic subsample was used in joint fits with the LAT data to test
models containing spectral breaks or cut-offs.

\section{Analysis} \label{sec:Analysis}

\subsection{LAT Upper Limits} \label{sec:LATUpperLimits}

We derive upper limits for the 288 GRBs that were detected by the GBM and fell in the LAT FOV from the LAT data using two methods.  The first consists of the standard unbinned likelihood analysis using the software developed and provided by the LAT team, while the second
method simply considers the total observed counts within an
energy-dependent acceptance cone centered on the GBM burst location.
The likelihood analysis will give more constraining upper limits, but
since it uses the instrumental point-spread-function (PSF) information to model the spatial
distribution of the observed photons, in cases where the burst
location is inaccurate and burst photons are present, it can give less
reliable constraints.  The latter method will be less constraining in
general, but it will also be less sensitive to errors in the burst
location, as the analysis considers photons collected over a fixed
aperture and does not otherwise use the burst or photon positions on
the sky.  We use both methods to obtain photon flux upper limits over a 0.1$-$10\,GeV energy range.

For the unbinned likelihood analysis, we used the standard software
package provided by the LAT team, (ScienceTools version v9r15p6)\footnote{{\tt http://fermi.gsfc.nasa.gov/ssc/}}.  We
selected ``transient'' class events in a $10^\circ$ acceptance cone
centered on the burst location, and we fit the data using the {\tt
  pyLikelihood} module and the {\tt P6\_V3\_TRANSIENT} response
functions \citep{Atwood:09}.  Each burst is modeled as a point source at the best
available location, derived either from an instrument with good localization capabilities (e$.$g$.$ \Swift\ or LAT) or by the GBM alone.  Of the 288 GRBs considered here, , In the likelihood fitting, the expected distribution of counts is modeled using the energy-dependent LAT PSF and a power-law source spectrum.
The photon index of the power-law is fixed to either the $\beta$ value
found from the fit of the GBM data for that burst or, if the GBM data
are not sufficiently constraining (i$.$e$.$ $\delta\beta \le 0.5$), to $\beta = -2.2$, the mean value
found for the population of BATSE-detected bursts \citep{Preece:00,Kaneko06}.
An isotropic background component is included in the model, and the
spectral properties of this component are derived using an empirical
background model \citep{GRB080825C} that is a function of the position
of the source in the sky and the position and orientation of the
spacecraft in orbit.  This background model accounts for contributions
from both residual charged particle backgrounds and the time-averaged
celestial gamma-ray emission.

Since we are considering cases where the burst flux in the LAT band
will be weak or zero, the maximum likelihood estimate of the source flux
may actually be negative owing to downward statistical fluctuations in
the background counts.  Because the unbinned likelihood function is
based on Poisson probabilities, a prior assumption is imposed that requires the
source flux to be non-negative.  This is necessary to avoid negative
probability densities that may arise for measured counts that are
found very close to the GRB point-source location because of the
sharpness of the PSF.  On average, this means that for half of the
cases in the null hypothesis (i.e., zero burst flux), the ``best-fit''
value of the source flux is zero but does not correspond to a local
maximum of the unconstrained likelihood function \citep{Mattox:96}.

Given the prior of the non-negative source flux, we treat the
resulting likelihood function as the posterior distribution of the
flux parameter.  In this case, an upper limit may be obtained by
finding the flux value at which the integral of the normalized
likelihood corresponds to the chosen confidence level
\citep{Amsler:08}.  For a fully Bayesian treatment, one would
integrate over the full posterior distribution, i.e., marginalize over
the other free parameters in the model.  However, in practice, we have
found it sufficient to treat the profile likelihood function as a
one-dimensional probability distribution function (pdf) in the flux parameter.  Again, in the limit of
Gaussian statistics and a strong source, this method is equivalent to
the use of the asymptotic standard error for defining confidence
intervals.  Hereafter, we will refer to this treatment as the
``unbinned likelihood'' method.

In the second set of upper limit calculations, we implement the method
described by \citet{Helene:83} and the interval calculation implemented in \citet{Kraft91}.  Here, the upper limit is computed in
terms of the number of counts and is based on the observed and
estimated background counts within a prescribed extraction region.
For the LAT data, the extraction region is an energy-dependent
acceptance cone centered on the burst position.  Since the burst
locations from the GBM data have typical systematic uncertainties
$\sim\,$3.2$^\circ$ (Connaughton et al. 2011), the size of the acceptance cone
at a given energy is taken to be the sum in quadrature of the LAT 95\% PSF
containment angle and the total (statistical + systematic) uncertainty
in the burst location.  The counts upper limits are evaluated over a
number of energy bands, converted to fluxes using the energy-dependent
LAT exposure at the burst location, and then summed to obtain the
final flux limit.  Since this method relies on comparing counts
without fitting any spectral shape parameters, we will refer to this
as the ``counting'' method.

The time intervals over which the upper limits are calculated are
important for their interpretation.  For both upper limit methods, we
consider three time intervals: two fixed intervals of 30 and 100
seconds post-trigger, and a ``T100'' interval that is determined
through the use of the Bayesian Blocks algorithm \citep{Jackson05} to estimate the
duration of burst activity in the NaI detector that has the largest
signal above background. For the T100 interval, an estimate of the
time-varying background count rate is obtained by fitting a 3rd degree
polynomial to the binned data in time intervals outside of the prompt
burst phase. Nominally, we take $\Tzero-dt$ to $\Tzero-100$\,s and
$\Tzero+150$\,s to $\Tzero+dt$, where $\Tzero$ is the GBM trigger time
and $dt = 200$\,s, although we increased the separation of these
intervals in some cases to accommodate longer bursts.  The counts per bin is then subtracted by the
resulting background model throughout the $\Tzero-dt$ to $\Tzero+dt$
interval, and the binned reconstruction mode of the Bayesian Blocks
algorithm is applied.  The T100 interval is then
defined by the first and last change points in the Bayesian Blocks
reconstruction.

The two fixed time intervals have been introduced so as to not bias
our results through assumptions regarding the durations of the high
energy components.  The brighter LAT-detected GRBs have exhibited
both delayed and extended high-energy emission on time scales that exceed
the durations traditionally defined by observations in the keV$-$MeV
energy range (Abdo et al. 2011).  Hence, we search for and place
limits on emission over intervals that may in some cases exceed the burst duration.  We will
discuss the implications of the limits found for the various time
intervals in section~\ref{sec:ResultsLATUpperLimits}.

\subsection{GBM Spectroscopy} \label{sec:GBMSpectroscopy}

For the 92 bursts in the bright BGO subsample, we performed spectral
fits to the NaI and BGO data and estimated the flux expected to
be seen by the LAT between $0.1$-$10$\,GeV using the GBM
fitted Band function \citep{Band93} parameters.  The selection of background and
source intervals for all bursts were performed manually through the
use of the RMFIT (version 3.3) spectral analysis software package\footnote{http://fermi.gsfc.nasa.gov/ssc/data/analysis/user/}.
Because the number of counts in the highest BGO energy bins is often
in the Poisson regime, we use the Castor modification \citep{Castor95} to the Cash statistic \citep{Cash76}, commonly referred to as C-Stat \footnote{{\tt http://heasarc.nasa.gov/xanadu/xspec/manual/manual.html}}, since the standard $\chi^2$ statistic is not reliable for low counts.  The variable GBM background for each burst is
determined for all detectors individually by fitting an
energy-dependent, second order polynomial to the data several hundred
seconds before and after the prompt GRB emission.  The standard 128
energy bin CSPEC data \citep{Meegan09} from the triggered NaI and BGO detectors were
then fit from 8\,keV to 1\,MeV and from 200\,keV to 40\,MeV,
respectively, for each burst.

As we noted above, only 30 bursts in the bright BGO subsample have
sufficient signal-to-noise to constrain the high-energy power-law
index $\beta$ of the Band function to within $\pm 0.5$.  Although we
considered a variety of models in our spectral analysis, we
found that the Band function was sufficient to describe
the spectral shape for all of these bursts. 

\section{Results} 
\label{sec:Results}

\subsection{LAT Upper Limits} 
\label{sec:ResultsLATUpperLimits}

Of the 288 GRBs in our sample, we were able to obtain upper limits, at
95\% confidence level (CL), for 270 bursts using the unbinned
likelihood method and 95\% CL upper limits for 250 bursts using the
counting method for the T100 intervals derived from the
GBM data.  The GRBs for which upper limits could not be calculated
were bursts that occurred either during spacecraft passages through
the South Atlantic Anomaly (SAA) or at angles with respect to the Earth's zenith that were $\ga
100^\circ$, thereby resulting in diffuse emission at the burst
locations that was dominated by $\gamma$-rays from the Earth's limb produced by interactions of cosmic rays with the earth's atmosphere.
These cases where the burst occurred at a high angle with respect to the zenith primarily affects the counting method,
because it requires a reliable estimate of the background during the
burst, and our method to estimate the background does not account for
Earth limb emission.  The likelihood method can fit for an Earth limb as a
diffuse component, but it may give weaker limits since the background
level is not as tightly constrained in this case compared to when the empirical
background estimate can be used to model all of the non-burst
emission.  The photon flux upper limits found for the likelihood
method for all three time intervals are presented in the last three
columns of Table~\ref{Table:SampleDefinition}.

\begin{figure*}
\includegraphics[width=7in,angle=0]{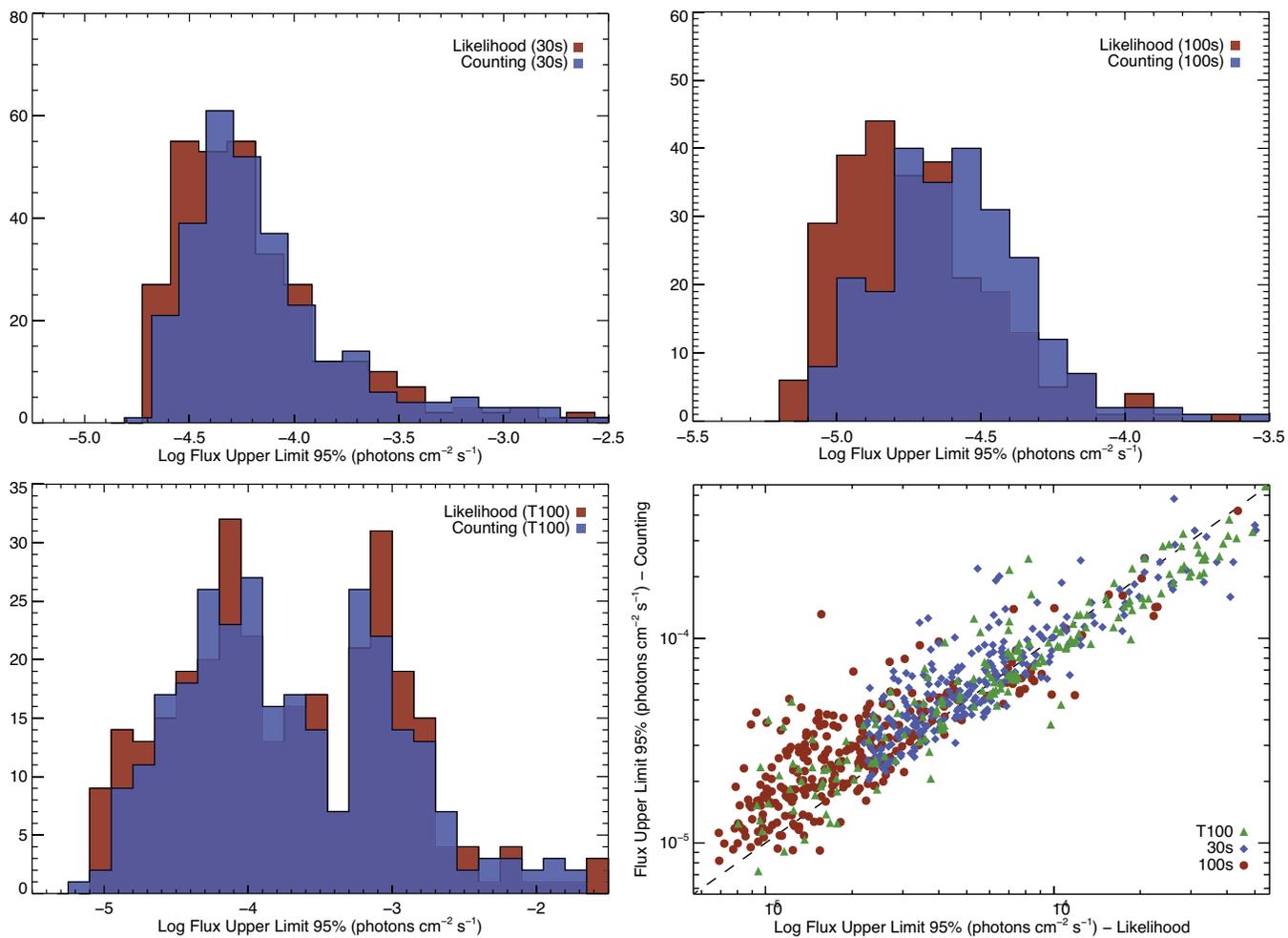}
\caption{The distributions of the 95\% CL photon flux upper limits obtained via the likelihood and counting methods for the 30\,s (upper-left), 100\,s (upper-right), and T100 (lower-left) time intervals.  A scatter plot comparison of the upper limits calculated over the three intervals is shown in the lower-right panel. The dashed line represents the line of equality between the likelihood and counting methods.}
\label{BayesianHeleneComparison}
\end{figure*}

The distributions of the 95\% CL photon flux upper limits obtained via the likelihood and
counting methods for the 30\,s, 100\,s, and T100 time intervals are
shown in upper-left, upper-right, and lower-left panels of Figure \ref{BayesianHeleneComparison}, respectively.  As expected, the
likelihood limits are systematically deeper than those found using the
counting method over the same time interval.  For either method, the upper limits for the
100\,s integrations are roughly half an order-of-magnitude deeper than
for the 30\,s integrations.  In the photon-limited case, this is
expected since the flux limit at a specified confidence level should
be inversely proportional to the exposure.  The doubly peaked upper limits distribution that appears in the upper-left panel of Figure \ref{BayesianHeleneComparison} 
for the T100 duration reflects the bimodal duration distribution for the short and long GRB populations.   The median of the T100 upper limit distribution for the likelihood method is
$\Fmed{T100} = 1.20\times10^{-4}$ photons cm$^{-2}$ s$^{-1}$ with a standard deviation of $\sigma_{T100} = 1.57\times10^{-3}$; whereas the counting method distribution has a median of $\Fmed{T100} = 1.27\times10^{-4}$ photons cm$^{-2}$ s$^{-1}$ and $\sigma_{T100} = 1.52\times10^{-3}$.  The median of the 30\,s upper limits distribution for the likelihood method is $\Fmed{30\rm s} = 4.76\times10^{-5}$ photons cm$^{-2}$ s$^{-1}$ with a standard deviation of $\sigma_{30\rm s} = 3.20\times10^{-4}$; whereas the counting method distribution has a median of $\Fmed{30\rm s} = 5.46\times10^{-5}$ photons cm$^{-2}$ s$^{-1}$ and $\sigma_{30\rm s} = 3.00\times10^{-4}$.  The median of the 100\,s upper limits distribution for the likelihood method  are $\Fmed{100\rm s} = 1.74\times10^{-5}$ photons cm$^{-2}$ s$^{-1}$ and $\sigma_{100\rm s} = 1.23\times10^{-4}$ and $\Fmed{100\rm s} = 2.59\times10^{-5}$ photons cm$^{-2}$ s$^{-1}$ and $\sigma_{100\rm s} = 1.06\times10^{-4}$ for the counting method.

\begin{figure*}
\includegraphics[width=7in,angle=0]{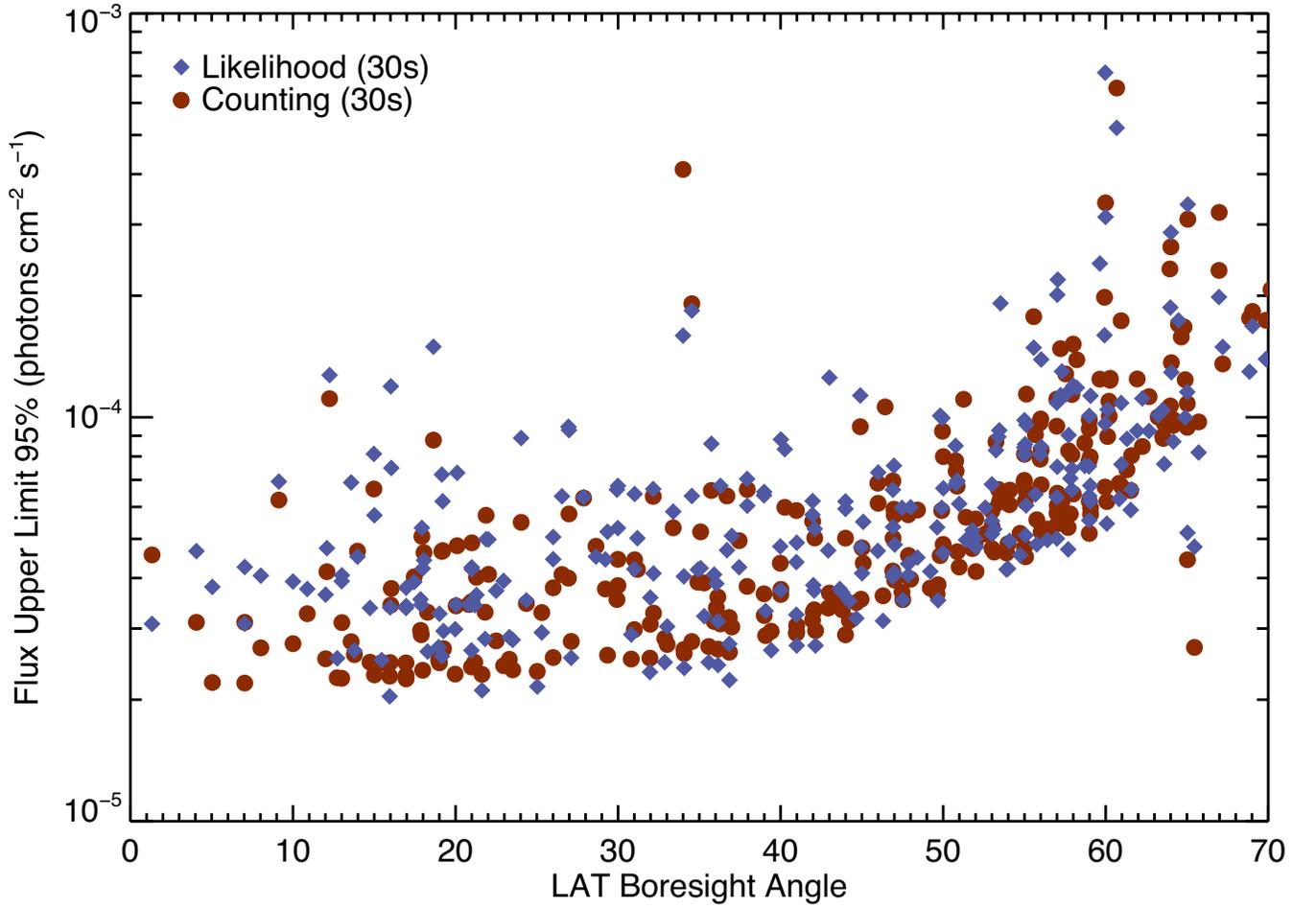}
\caption{The 95\% CL photon flux upper limits determined using the likelihood and counting methods as a function of off-axis angle.  The decreasing exposure as a function of off-axis angle results in the shallowing of the LAT upper limits for bursts occurring away from the LAT bore sight.}
\label{BayesianHeleneVsAngle}
\end{figure*}

A comparison of the likelihood and counting methods for all three time intervals for is shown in the lower-right panel of Figure~\ref{BayesianHeleneComparison}.  The scatter in the upper limits distribution for both methods is largely due to the
range of angles at which the GRBs occurred with respect to the LAT
boresight, resulting in different effective areas and hence different
exposures for each burst.  The LAT exposure as a function of the off-axis angle drops steeply with increasing inclination, resulting in a shallowing of the LAT upper limits as a function of increasing off-axis angle, which can be seen in Figure~\ref{BayesianHeleneVsAngle}.  
Overall, the two methods give consistent results for the bursts in our sample, and
therefore we will hereafter focus primarily on the limits obtained with the
likelihood method in our discussion of the implication of these
results.

\newcommand{\phcms}{photons\,cm$^{-2}$s$^{-1}$}

Despite the dependence of the upper limit values on off-axis angle,
the distribution of LAT photon flux upper limits is relatively narrow
for angles $< 40^\circ$, allowing us to define an effective LAT
sensitivity assuming a typical GRB spectrum (i.e., $\beta \approx
-2.2$).  We can therefore set sensitivity thresholds for the corresponding median photon flux upper limit for each integration time of
$F_{\rm lim,30s} = 4.7\times10^{-5}$\,\phcms and $F_{\rm lim,100s} = 1.6\times10^{-5}$\,\phcms.  

\begin{figure*}
\includegraphics[width=7in,angle=0]{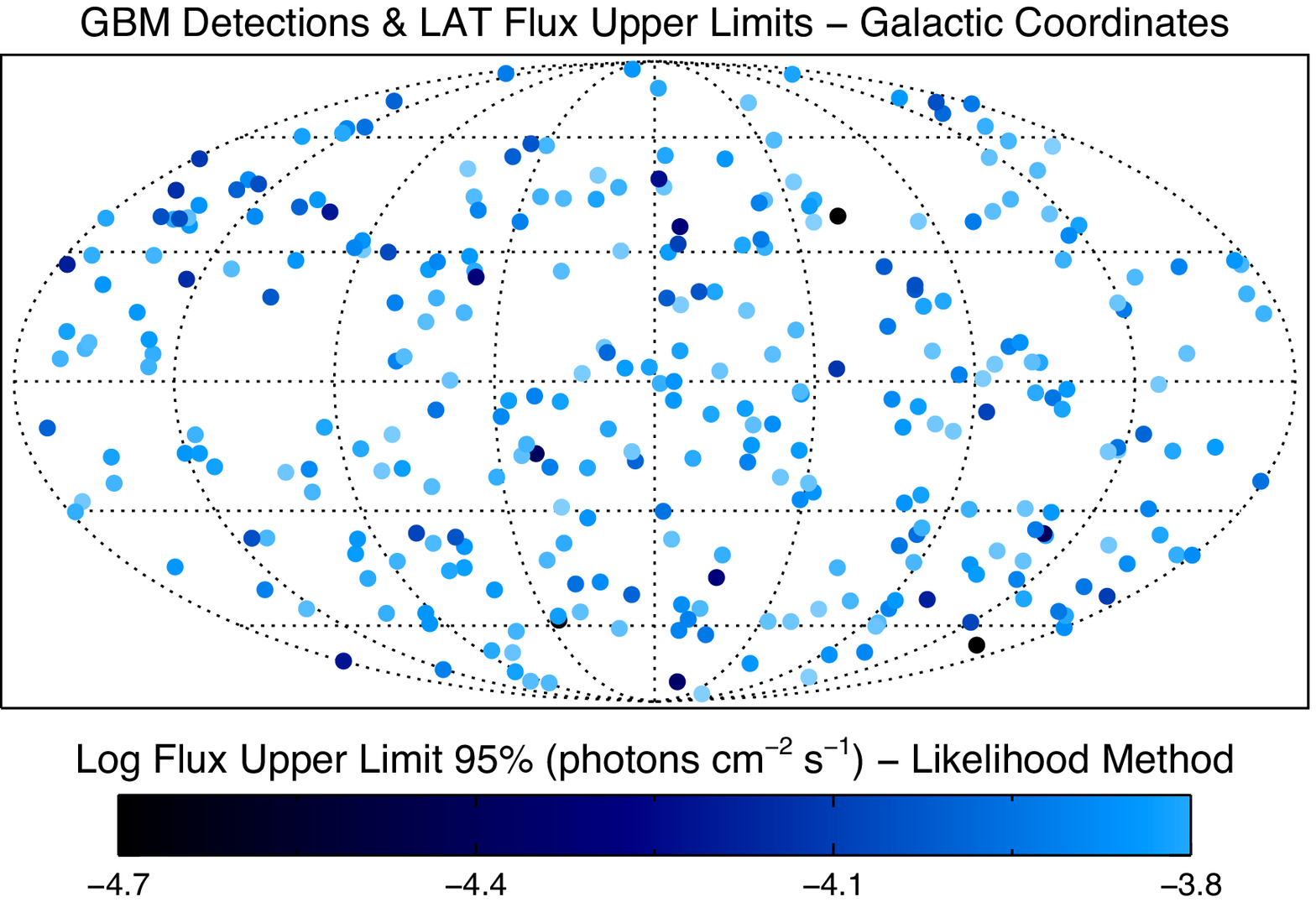}
\caption{The celestial distribution of 288 gamma-ray bursts as detected by \Fermi-GBM in the first 2.5 years of LAT operations that fell in the LAT FOV, plotted in Galactic coordinates.  The colors represents the 95\% CL LAT photon flux upper limits.}
\label{SkyMap}
\end{figure*}

Finally, in Figure~\ref{SkyMap} we plot the location of each burst on
the sky in Galactic coordinates, color-coded to represent the
likelihood-determined photon flux upper limits.  There is no evidence
of a spatial dependence of the GBM detection rate nor of the magnitude
of the LAT upper limit, as a function of Galactic latitude $b$.  

\subsection{GBM Spectral Fits and Upper Limit Comparisons} 

We compare the LAT upper limits calculated over the burst duration to the expected 0.1$-$10\,GeV photon fluxes found through extrapolations of spectral fits presented in the first GBM spectral catalog (Goldstein et al$.$ in press).  We focus this analysis on bursts for which a Band spectral model was a preferred fit compared to models with fewer degrees of freedom, since alternative models such as Comptonized spectra suffer sharp drops in expected flux at high-energy and are not expected to result in LAT detections without the presence of additional spectral components.  Of the 487 GRBs presented in that catalog, a Band model fit was preferred over simpler models for 161 bursts, 75 of which appeared in the LAT field of view.  For this comparison, the LAT upper limits were recalculated for a duration that matched the interval used in the GBM spectral catalog (see Goldstein et al. 2011 for a detailed discussion of their interval selection).  We next performed a simulation in which we varied the expected LAT photon flux fitted values using the associated errors for each burst in order to determine the median number of bursts over all realizations that would fall above the LAT upper limit.  In a total of $10^{5}$ realizations, we find that 50$\%$ of the GRBs in the GBM spectral catalog, which prefer a Band model fit, have expected 0.1$-$10\,GeV photon fluxes that exceeds the LAT upper limit.

We investigate the differences between the GBM-based extrapolations and the LAT upper-limits further by performing detailed spectral fits to our spectroscopic subsample.  The spectral parameters obtained from the fits to the GBM data only for the 30 GRBs in this spectroscopic subsample are listed in
Table~\ref{Table:SpectralParameters}.  The median values of the low and high
energy power-law indices and the peak of the $\nu F_{\nu}$ spectra are
$\alpha = -0.83$, $\beta = -2.26$, and $\Epeak = 164\,$keV, with
standard deviations of $\sigma_\alpha = 0.44$, $\sigma_\beta = 0.25$,
and $\sigma_{\Epeak} = 177$\,keV, respectively.  The distributions
of spectral parameters for these bursts are consistent with similar
distributions found for BATSE-detected GRBs \citep{Preece:00,Kaneko06}. The
time durations used in the spectral fits and the
time-averaged photon flux values in the 0.02$-$20\,MeV energy range for
these GRBs are given in Table~\ref{Table:ExpectedFlux}.  In the third
column, we list the expected flux in the 0.1$-$10\,GeV energy range
assuming a power-law extrapolation of the Band function fit to the
GBM data; and in the fourth column, we give the measured LAT photon
flux upper limit found for the same time interval.  The errors on the
expected LAT photon fluxes were determined using the covariance
matrices obtained from the GBM spectral fits.

\begin{figure*}
\includegraphics[width=7in,angle=0]{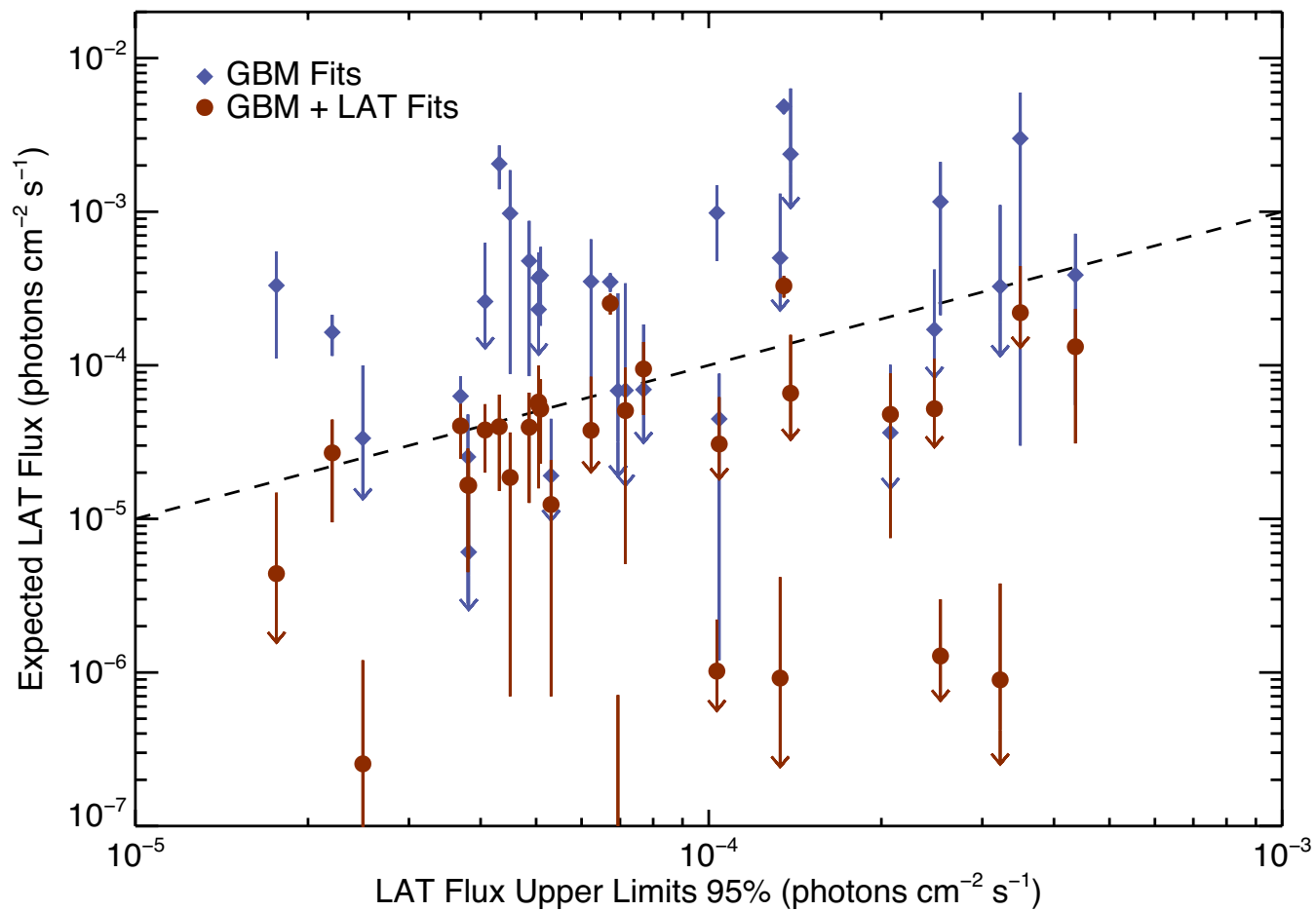}
\caption{The expected photon flux, based on fits to the prompt GBM spectrum and duration plotted versus the LAT flux upper limit for each burst.  When fitting only to the GBM data, roughly 50$\%$ of the bursts in the spectroscopic sample have expected LAT fluxes that exceed the LAT 95\% CL flux upper limit.  When fitting both the GBM and LAT data, only 23\% of our sample have expected flux values that exceed the 95\% CL LAT flux upper limit. The dashed line represents the line of equality.}
\label{ExpectedFluxComaprison_SpectroscopicDuration}
\end{figure*}

\begin{figure*}
\includegraphics[width=7in,angle=0]{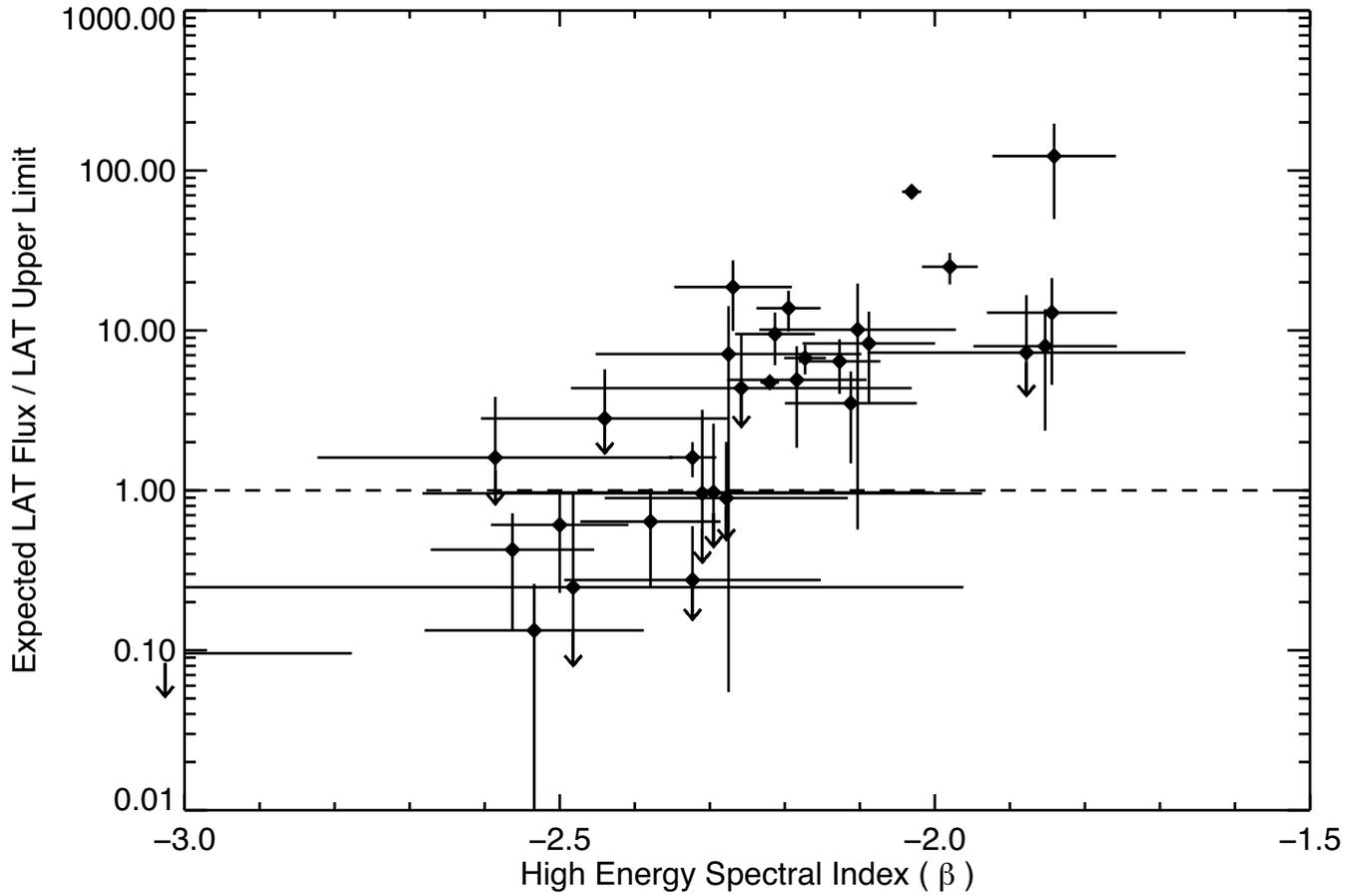}
\caption{The ratio of the expected LAT flux, based on fits to the prompt GBM spectrum, to the LAT 95\% CL LAT flux upper limit plotted versus the GBM determined high-energy spectral index.  The degree to which the expected flux in the LAT energy range from these bursts exceed our estimated LAT upper limits correlates strongly with the measured high-energy spectral index.}
\label{BetaVsRatio}
\end{figure*}

A comparison of the LAT photon flux upper limits versus the expected 0.1$-$10\,GeV photon fluxes for
each burst in our spectroscopic subsample is shown as blue data points in Figure~\ref{ExpectedFluxComaprison_SpectroscopicDuration}.  The
downward arrows on the expected flux values indicate values that are
consistent with zero within the 1-$\sigma$ errors shown.  The dashed
line represents the line of equality between the expected LAT
photon flux and the LAT photon flux upper limits when calculated for the durations presented in Table~\ref{ExpectedFluxComaprison_SpectroscopicDuration}.  In a total of $10^{5}$ realizations, we find that 53$\%$ of GRBs in our spectroscopic subsample have expected 0.1$-$10\,GeV photon fluxes that exceed their associated 95\% CL LAT upper limit.  As with the flux comparison, roughly 50$\%$ in our sample also have expected
fluence values that exceed the 95\% CL LAT fluence upper limit.
Figure~\ref{BetaVsRatio} shows that the degree to which the expected
flux in the LAT energy range from these bursts exceed our estimated
LAT upper limits correlates strongly with the measured high energy
spectral index, with particularly hard bursts exceeding the estimated
LAT sensitivity by as much as a factor of 100.  Again, the spectral fits to the bright bursts detected by the BGO clearly shows that a simple extrapolation from the GBM band to the LAT band systematically over-predicts the observed flux.

\subsection{Joint GBM and LAT Spectral Fits} 
\label{sec:Joint_GBM_LAT_Fits_I}

Including the LAT data in the spectral fits drastically alters the
best-fit Band model parameters and the resulting expected photon flux
in the LAT energy range.  The best-fit parameters of the joint
spectral fits for the spectroscopic subsample can be found in
Table~\ref{Table:GBMLAT_SpectralFitParameters}.  The high-energy spectral indices are
typically steeper (softer) than found from fits to the GBM data alone.

The difference in the $\beta$ values for the joint fits with respect
to the fits to the GBM data alone can be found in Column 8 of
Table~\ref{Table:GBMLAT_SpectralFitParameters}.  The resulting $\beta$ distributions are
shown in Figure~\ref{BetaGBMLATComparison_Combined}.  The GBM-only
$\beta$ distribution (red histogram) peaks at $\beta = -2.2$, matching
the $\beta$ distribution found for the population of BATSE-detected
bursts presented in \citet{Preece:00}.  In contrast, the $\beta$
distribution found from the joint fits (blue histogram) indicates
spectra that are considerably softer, with a median value of $\beta =
-2.5$.  While the GBM-only $\beta$ distribution includes 5 GRBs with $\beta > -2.0$, no bursts
had $\beta$ values this hard from the joint fits.  The low energy power-law index $\alpha$ and the peak of the
${\nu}F_{\nu}$ spectra, $\Epeak$ distribution remain relatively
unchanged.  In Figure \ref{ExpectedFluxComaprison_SpectroscopicDuration}, we compare the LAT photon flux upper limits calculated over the
burst duration presented in Table~\ref{Table:GBMLAT_SpectralFitParameters} versus the expected 0.1$-$10\,GeV photon fluxes for
each burst, now using a power-law extrapolation of the Band function that was fit to both
the GBM and LAT data.  The softer $\beta$ values
obtained through the joint fits yield expected LAT photon flux values
that are more consistent with the LAT non-detections, with only 23$\%$ of the bursts in our spectroscopic subsample with expected
flux values that exceed the 95\% CL LAT flux upper limit given  $10^{5}$ realizations of the data about their errors.  We find that a similar ratio of bursts have expected fluence values that exceed their associated 95\% CL LAT fluence upper limit.

\subsection{Spectral Breaks or Softer Spectral Indices?}

Although the discrepancy between the predicted 0.1$-$10\,GeV fluxes
from the GBM-only fits and the LAT upper limits can be explained by the
softer $\beta$ values in the joint fits, intrinsic spectral breaks at
energies $\ga 40$\,MeV can also reconcile the conflicting GBM and LAT
results. Determining whether softer $\beta$ values or spectral breaks
are present has at least two important implications: If the spectral
breaks or cut-offs arise from intrinsic pair production ($\gamma\gamma \to e^+e^-$) in the source, then the
break or cut-off energy would provide a direct estimate of the bulk
Lorentz factor of the emitting region within the outflow
On the other hand, an intrinsically softer distribution of $\beta$
values would mean that theoretical inferences based on the $\beta$
distributions found by fitting BATSE or GBM data alone may need to be
revised.  Evidence for either spectral breaks or softer $\beta$ values could also provide support for multi-component models that have been used to describe novel spectral features detected by the GBM and LAT \citep[e$.$g$.$,][]{Guiriec2011}.

For the joint fitting of the GBM and LAT data, deciding between the
two possibilities for any single burst can be cast as a standard model
selection problem.  Under the null hypothesis, we model the GRB
spectrum using a simple Band function, as we have done in
section~\ref{sec:Joint_GBM_LAT_Fits_I}.  As an alternative hypothesis,
we could extend the Band model to account for the presence of a
spectral break.  This may be done via an additional break energy above
the Band $\Epeak$, effectively using a doubly broken power-law in the
fit; or it could be accomplished by adding an exponential cut-off to
the Band model with cut-off energy $E_c > \Epeak$.  In either case, the
null and alternative hypotheses are ``nested'' such that the former is
a special case of the latter for some values of the extra model
parameters that are introduced.  Assuming there are $n_{\rm alt}$
additional free parameters under the alternative model, then whether
the alternative model is statistically preferred would be given by the
$\Delta$C-Stat value assuming it follows a $\chi^2$ distribution
for $n_{\rm alt}$ degrees-of-freedom.

For the purposes of this analysis, we have adopted an alternative
model consisting of a Band function plus a step function fixed at
50\,MeV.  Although unphysical, a simple step function introduces
a single additional degree-of-freedom and can adequately represent the
need for a break in the high-energy spectra.  This additional degree-of-freedom represents the normalization of the Band function's high-energy component above 50\,MeV, which is left to vary, leading to the normalization of the power-law above 50 MeV being adjusted such that it is always consistent with the LAT upper limits.  For this analysis, the index of the power-law above the break is fixed to match the Band function's high-energy power-law index, which is allowed to vary as a free parameter. Since this introduces a single extra degree-of-freedom, a value of $\Delta$C-Stat $> 9$ would represent a >$3\sigma$ improvement in the fit.  We adopt this criterion as the threshold for a statistical preference for a break in the high-energy spectrum of an individual GRB.

\begin{figure*}
\includegraphics[width=7in, angle=0]{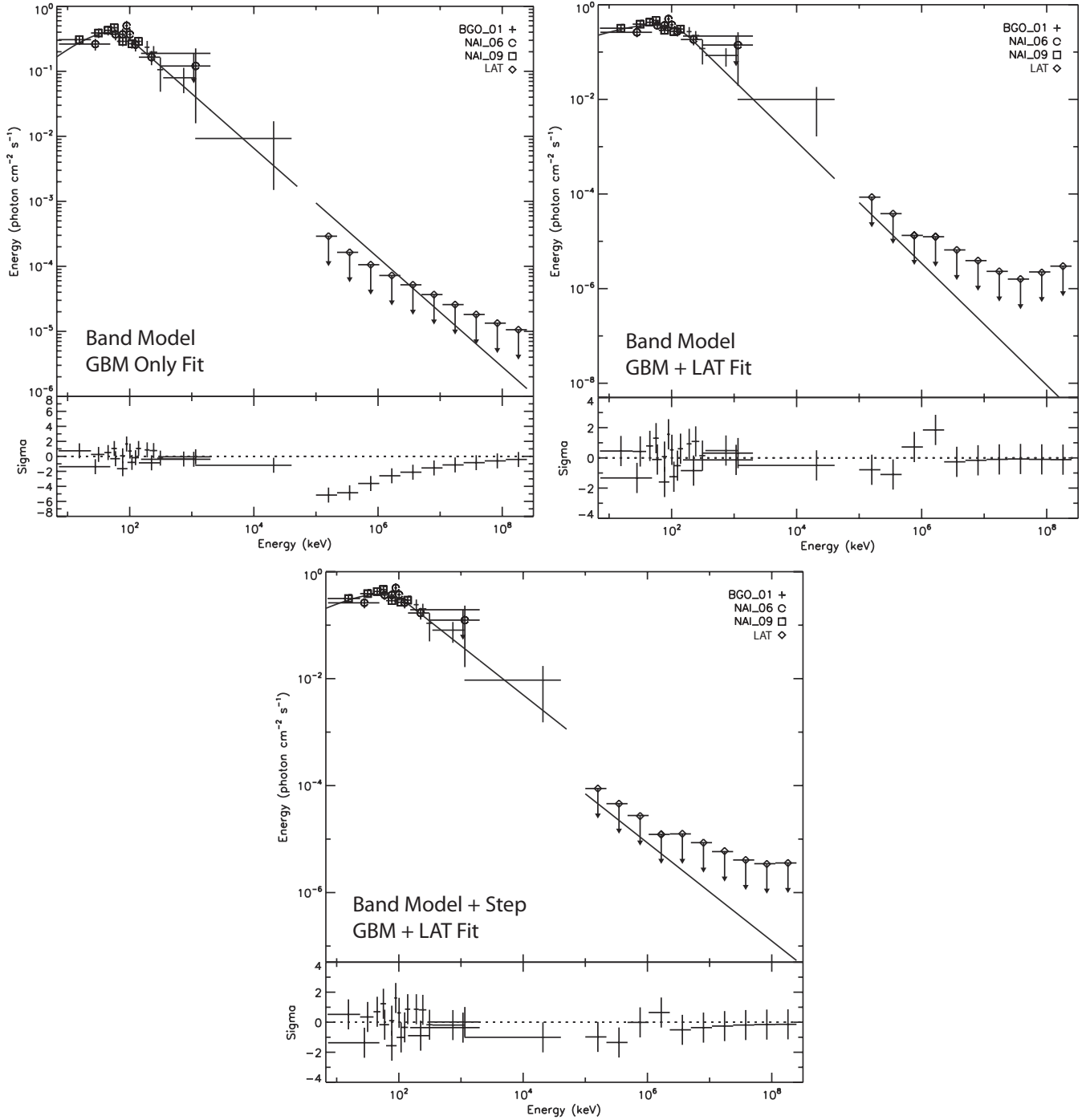}
\caption{Example spectral fits showing (clockwise) a Band model fit to GBM data alone, a Band model fit to both the GBM and LAT data, and a Band model plus a step function fit to the GBM and LAT data.}
\label{SpectralFitExample}
\end{figure*}

An example of such a fit can be seen in
Figure~\ref{SpectralFitExample}, where the three panels show (clockwise) a Band model fit to GBM data alone, a Band model fit to
both the GBM and LAT data, and a Band model plus a step function fit
to the GBM and LAT data.  The difference between the first two panels
demonstrates the degree to which the high-energy spectral index can
steepen to accommodate the LAT data, despite being outside of the
range allowed by the statistical uncertainty in the $\beta$
determination made through the GBM fit alone.  The third panel shows
the effect of introducing a step function between the two instruments,
in which the requirement for a softer $\beta$ value is
alleviated.  For the fit shown in Figure \ref{SpectralFitExample}, the $\beta$ value determined through the Band model plus a step function fit is consistent with the value found by fitting a Band model to the GBM data alone.  

\begin{figure*}
\includegraphics[width=7in,angle=0]{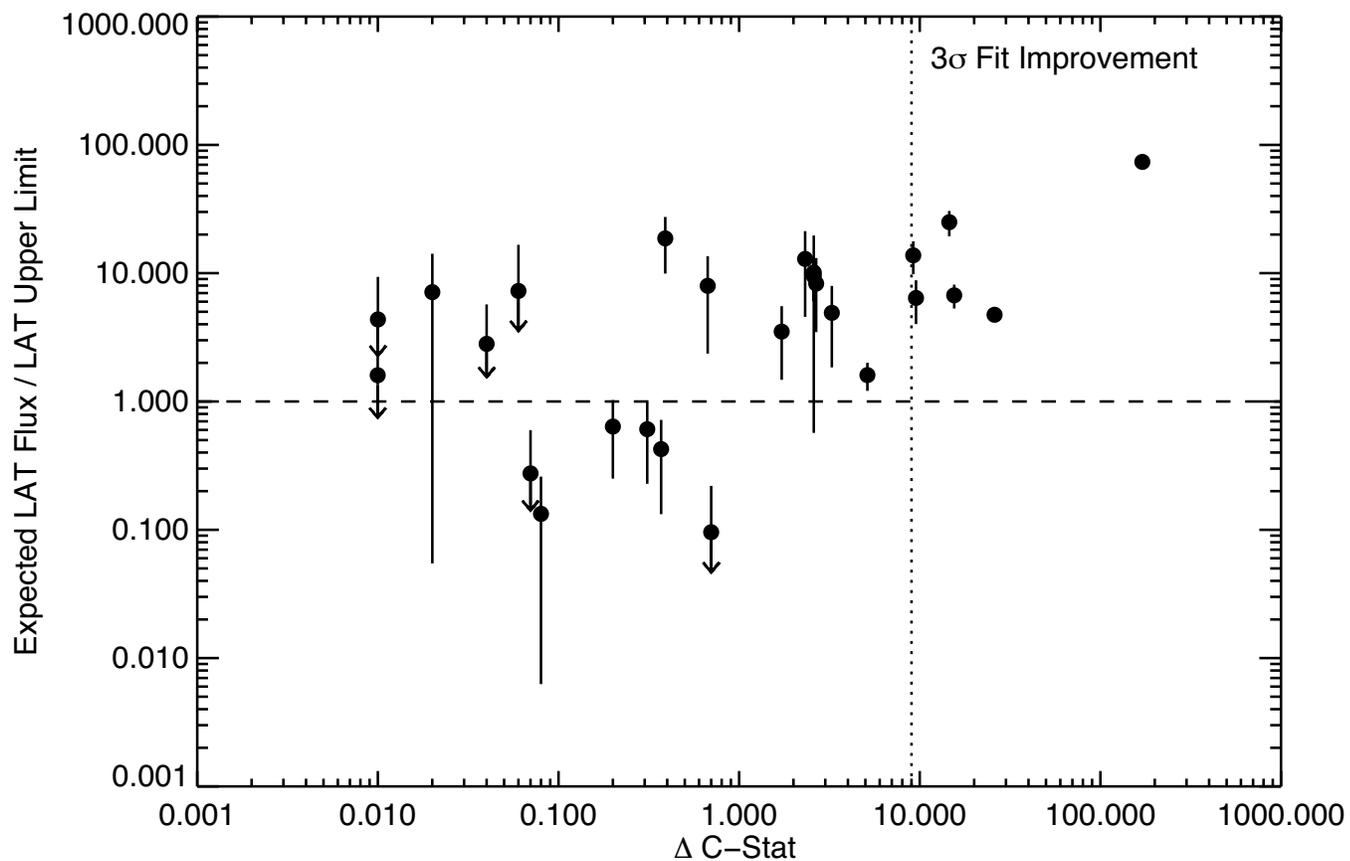}
\caption{The ratio of the expected LAT flux (based on GBM-only fits) to the LAT 95\% CL upper limit versus the $\Delta$C-Stat values for our spectroscopic subsample.  The long and short dashed lines represents the line of equality between the LAT upper limits and the expected LAT flux and the $\Delta$C-Stat value representing a $3\sigma$ fit improvement respectively.  The bursts for which a spectral break is statistically preferred have the most severe discrepancies between the GBM-only extrapolations and the LAT upper limits.}
\label{deltacstat}
\end{figure*}

\begin{figure*}
\includegraphics[width=7in,angle=0]{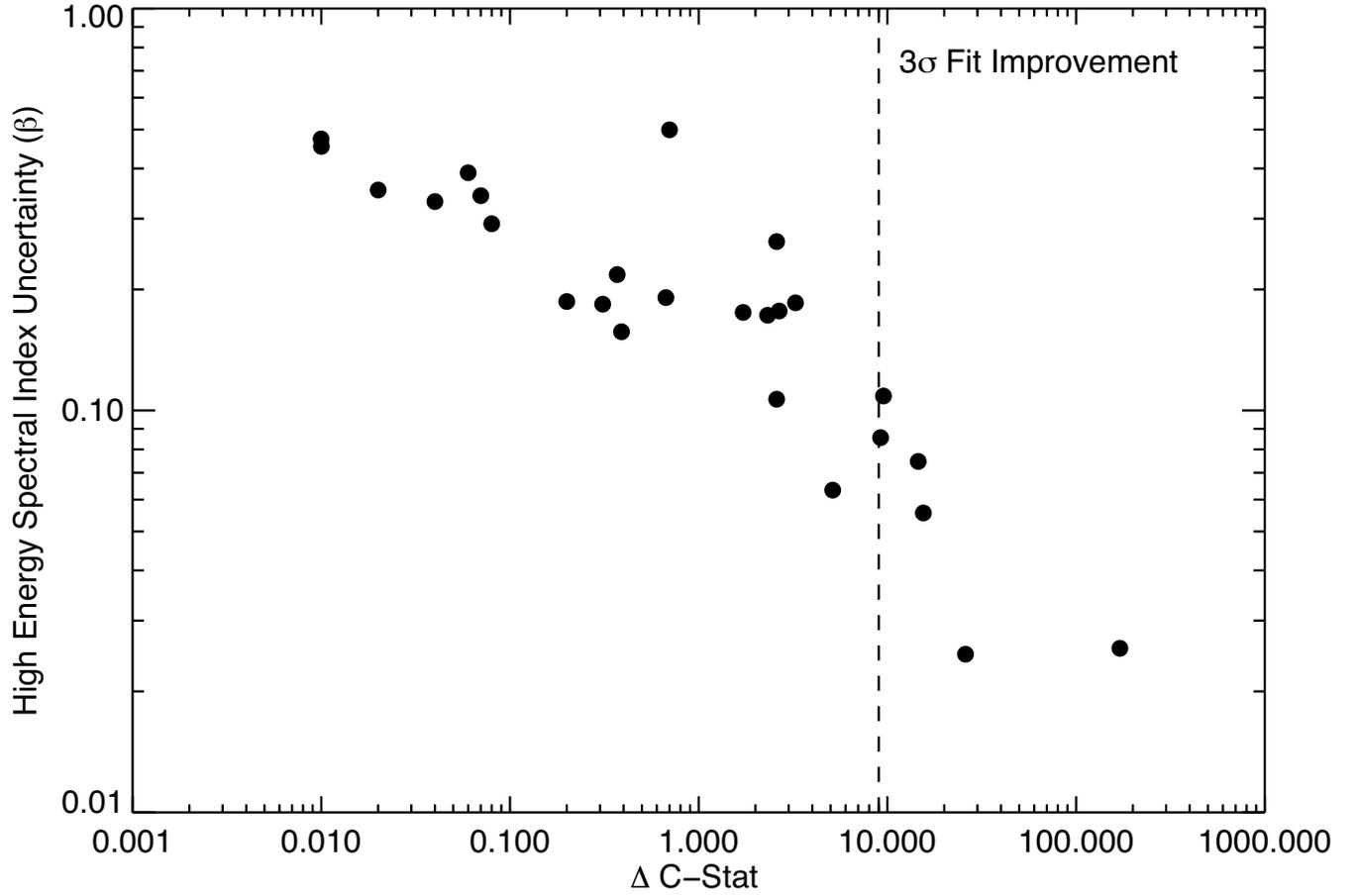}
\caption{The $1\sigma$ symmetric uncertainty in the high-energy spectral index found from fits to the GBM data alone versus the $\Delta$C-Stat values for our spectroscopic subsample.  The bursts for which a spectral break is statistically preferred also have the smallest uncertainties in their GBM-only $\beta$ values. }
\label{DeltaCStatVsBetaError}
\end{figure*}

The $\Delta$C-Stat values obtained for the Band and Band+step
function fits are listed in Column 9 of Table~\ref{Table:GBMLAT_SpectralFitParameters}.
For most of the bursts, a simple steepening of the high energy
power-law index was sufficient to explain the lack of a LAT detection.
However, in 6 cases $\Delta$C-Stat exceeded a value of
9, indicating a statistical preference for a break in the high
energy spectrum.  Figure~\ref{deltacstat} shows the ratio of the
expected LAT flux (based on GBM-only fits) to the LAT 95\% CL upper
limit plotted versus the $\Delta$C-Stat values for the spectroscopic
subsample.  A weak correlation between the flux ratio and $\Delta$C-Stat is
apparent.  In addition, Figure~\ref{DeltaCStatVsBetaError} shows an
anti-correlation between the resulting $\Delta$C-Stat values for this
sample plotted versus the uncertainty in the high-energy spectral
index found from fits to the GBM data alone.  The bursts for which a spectral break is statistically preferred both
have the most severe discrepancies between the GBM-only extrapolations
and the LAT upper limits and also have the smallest uncertainties
in their GBM-only $\beta$ values. 

\subsection{Constraints on the Bulk Lorentz Factor}

If we assume that the high-energy spectra in the 6 GRBs that prefer spectral cut-offs are a result of $\gamma\gamma$ attenuation, as opposed to a spectral turnover that is intrinsic to the GRB spectrum, then we can use the joint GBM and LAT spectral fits in conjunction with the LAT non-detections at 100 MeV to place limits on the maximum Lorentz factor.  In this context, the high-energy $\gamma$-rays produced within the GRB jet may undergo $\gamma\gamma\to e^+e^-$ pair production and can be
absorbed {\em in situ}.  The interaction rate of this process and
corresponding optical depth, $\tau_{\gamma\gamma}$, depend on the
target photon density and can be significant when both the high-energy
and target photons are produced in the same physical region.  Highly
relativistic bulk motion of such an emission region can reduce the
implied $\gamma\gamma$ optical depth greatly by allowing for a larger
emitting region radius and a smaller target photon density for a given
observed flux and variability time scale.  Observation of $\gamma$-ray
emission up to an energy $E_\max \gg m_e c^2$ thus can be used to
put a lower limit on the bulk Lorentz factor $\Gamma$ of the emitting
region~\citep{ls01,rmz04,gcd08,GRB090510_physics}.  This method is
valid for $\Gamma \le E_\max(1+z)/m_ec^2$, which follows from the
threshold condition for $e^+e^-$ pair production, when both the
incident and target photons are at the maximum observed energy.

If a high-energy $\gamma$-ray photon with energy $E$ and the observed
broadband photon emission originate from the same physical region, and if we assume the photons are quasi-isotropic in the comoving frame, then the $\gamma\gamma\to e^+e^-$ pair production optical depth can be
written as
\begin{eqnarray}
\tau_{\gamma\gamma}(E) = \frac{3}{4} \frac{\sigma_T d_L^2}{t_v\Gamma}
\frac{m_e^4c^6}{E^2(1+z)^3} 
\int_{\frac{m_e^2c^4\Gamma}{E(1+z)}}^{\infty}
\frac{d\epsilon^\prime}{\epsilon^{\prime 2}}
~n\left( \frac{\epsilon^\prime\Gamma}{1+z} \right)
\varphi\left[ \frac{\epsilon^\prime E(1+z)}{\Gamma} \right].
\label{gg_opacity}
\end{eqnarray}
Here $n(\epsilon)$ is the observed photon spectrum, $\epsilon$ is the
target photon energy, $\epsilon^\prime$ is the target photon energy in
the comoving frame of the emitting plasma, $d_L$ is the luminosity
distance, $t_v$ is the $\gamma$-ray flux variability time scale, and
$\sigma_T$ is the Thomson cross-section.  The function
$\varphi[\epsilon^\prime E(1+z)/\Gamma]$ is defined by \citet{gs67}
and \citet{bmg73}.  The value of $\Gamma_{\gamma\gamma,\min}$ follows from the
condition $\tau_{\gamma\gamma}(E_\max) = 1$.  This single-zone model, in which the spatial and temporal dependancies of $\tau_{\gamma\gamma}$ have been averaged out, has been the technique used to measure the reported values of $\Gamma_{\gamma\gamma,\min}$ for the LAT detections of GRBs 080916C, 090510, and 09092B in \citet{080916C}, \citet{GRB090510_physics}, and \citet{090902B}, respectively.

A direct estimate of the bulk Lorentz factor $\Gamma$, as opposed to a minimum value, of the GRB jet
can be made based on evidence of a cut-off in the spectral fits that
are attributed to $\gamma\gamma$ attenuation, such as has been reported for GRB~090926A in \citet{grb090926}. 

In the case of the 6 GRBs that we are considering here for which no direct evidence for a spectral cut-off is otherwise detected, we use our upper limits to calculate a maximum bulk Lorentz factor $\Gamma_{\gamma\gamma,\max}$ from the condition $\tau_{\gamma\gamma}(E_{\rm UL}) = 1$.  To do so, we use the Band function fit to the GBM and LAT data and set $E_{\rm UL} = 100$ MeV. We also assume a variability time scale of $t_v = 0.1$\,s, which we believe represents a conservative estimate of $t_v$ given the ubiquity of millisecond variability in BATSE detected GRBs \citep{Walker00} as well as the short timescales observed in other LAT detected GRBs \citep{GRB090510_physics}.  

We note that if the cutoff energy due to intrinsic pair opacity is small enough, $E_{\rm cutoff} < m_e c^2\Gamma/(1+z)$, then the Thomson optical depth of the pairs that are produced in the emitting region is $\tau_{T,e^\pm} > 1$ \citep{ls01,Abdo09}.  This should affect both the observed spectrum, thermalizing it for a large enough optical depth, and light curve, eliminating short timescale variability. For $E_{\rm cutoff} = 100\;$MeV, this condition is nearly violated at $z \lesssim 1.0$, therefore a much lower cutoff energy would be hard to reconcile with an intrinsic pair opacity origin for GRBs at low redshift.

\begin{figure*}
\includegraphics[width=7in,angle=0]{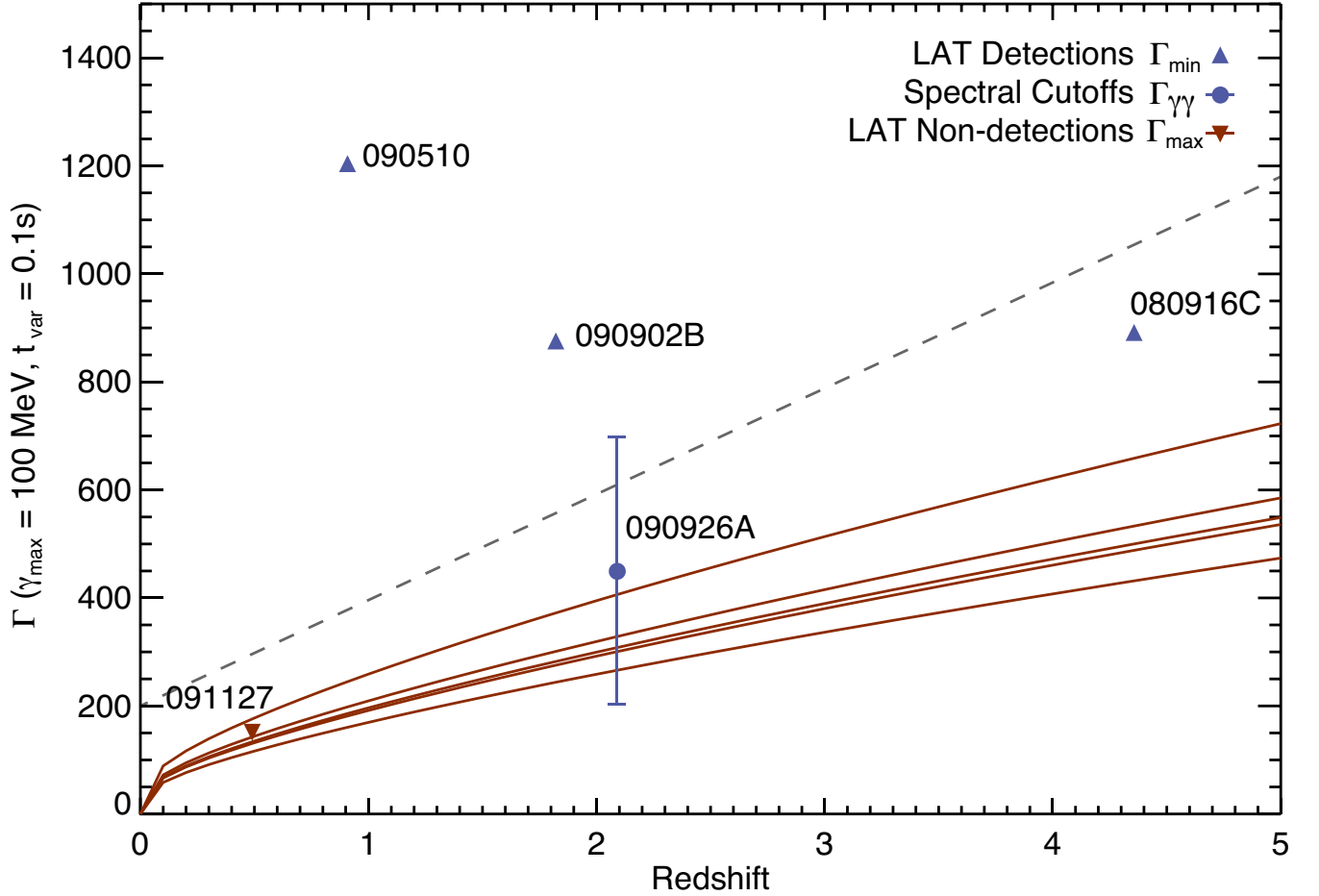}
\caption{The $\Gamma_\max$ values for the 6 GRBs in our sample with evidence for spectral breaks compared to the $\Gamma_\min$ values for the brightest LAT-detected GRBs.  The allowed range of $\Gamma_\max$ values for $0 < z < 5$ all lie well below the $\Gamma_\min$ values of the LAT-detected GRBs.  The $\Gamma$ estimate for GRB~090926A from \citet{grb090926} is shown as the filled blue circle.   The grey dashed line demarcates the self-consistency line where the condition that $\Gamma \le E_\max(1+z)/m_ec^2$ is violated.  The range of Lorentz factors obtained through the use of single-zone and time-dependent models places GRB~090926A between the LAT detected and LAT dark GRBs.}
\label{LorentzFactorVsRedshift}
\end{figure*}

The resulting $\Gamma_{\gamma\gamma,\min}$ and $\Gamma_{\gamma\gamma,\max}$ values for previously reported LAT detections and from the upper limits presented here are shown in Figure \ref{LorentzFactorVsRedshift}.  Since the Lorentz factor calculation depends on the redshift, which is
unknown for the majority of GBM detected bursts, we have plotted the
$\Gamma_{\gamma\gamma,\max}$ values as a function of the redshift (red lines).
One GRB in our spectroscopic subsample, GRB~091127, has a measured redshift which allows us
to constrain the burst's $\Gamma_\max$ value.  Using a redshift
of $z = 0.490$  \citep{Cucchiara09} and $E_{\rm UL} \sim 100$ MeV, we calculate a
relatively small bulk Lorentz factor of $\Gamma_\max \sim 155$.  Using the measurements of $E_{\rm UL}$ for these GRBs
provides a relatively narrow distribution of $\Gamma_\max$ that
range from $50 < \Gamma_\max < 300$ at $z = 1$ to $400 <
\Gamma_{\gamma\gamma,\max} < 640$ at $z = 4$.  These values stand in stark contrast to the LAT detected GRBs for which $\Gamma_{\gamma\gamma,\min}$ was measured, all of which have $\Gamma_{\gamma\gamma,\min} > 800$.  

The detection of spectral curvature by the LAT in the spectrum of GRB~090926 provides a case that appears to bridge the LAT detected and non-detected samples.  The estimate of $\Gamma$ of 200$-$700 presented in \citet{grb090926} reflects the systematic differences between Lorentz factors obtained through the use of time-dependent models by \citet{gcd08} which yield systematic differences in $\tau_{\gamma\gamma}$ and the inferred $\Gamma$ when compared to the simple single-zone model used above.  \citet{gcd08}, and more recently \citet{HD11}, have shown that such time-dependent models, which include the temporal evolution of $\tau_{\gamma\gamma}$ during the emission period, can yield inferred $\Gamma$ estimates that are reduced by a factor of 2-3 compared to estimates made using single-zone models.  In the context of these time-dependent model, the $\Gamma_{\gamma\gamma,\min}$ and $\Gamma_{\gamma\gamma,\max}$ presented in Figure \ref{LorentzFactorVsRedshift} would all be systematically overestimated by a factor of 2-3, but the dichotomy between the LAT detected and LAT non-detected GRBs would persist since all $\Gamma$ estimates would be effected by the same correction.

Note that the grey dashed line in Figure \ref{LorentzFactorVsRedshift} demarcates the self-consistency line where the condition that $\Gamma \le E_\max(1+z)/m_ec^2$ is violated, implying an incorrect determination of $\tau_{\gamma\gamma}$, for the bursts with no detected emission above $E_\max = 100$ MeV.  None of the bursts in our spectroscopic subsample violate this condition at any redshift for the choice of $E_{\rm cutoff} = 100\;$MeV.

\section{Discussion} \label{sec:Discussion}

The upper limits presented above place stringent constraints on the high
energy emission from GRBs detected by the GBM.  Of the 620 bursts
detected by the GBM from 2008 August 4th to 2011 January 1st, 46\% were
within the LAT FOV.  There is evidence for high energy
emission $> 100$ MeV in the LAT energy range for 23 GRBs, representing 8\% of
the entire GBM sample observed by the LAT.  This is significantly less
than the pre-launch estimate of 1 detection per month
that produces at least 100 counts above 100 MeV \citep{Band09}.

\begin{figure*}
\includegraphics[width=7in,angle=0]{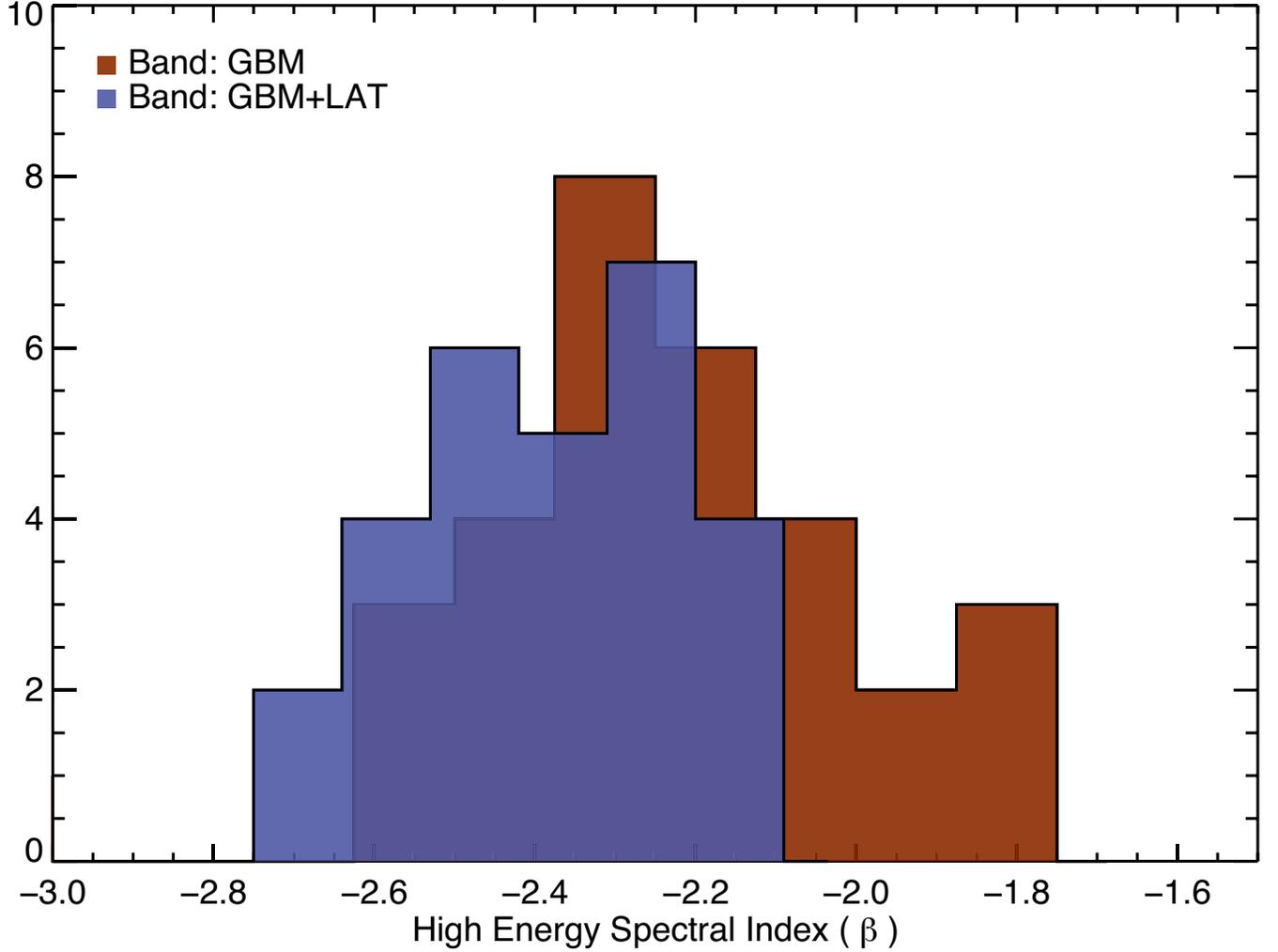}
\caption{A comparison between the high-energy spectral indices measured through spectral fits to the GBM data alone and joint fits to both the GBM and LAT data.   The GBM-only
$\beta$ distribution has a median value of $\beta = -2.2$, matching the distribution found by \citep{Preece:00,Kaneko06}.  In contrast, the $\beta$ distribution found from the joint fits indicate spectra that are considerably softer, with a median value of $\beta = -2.5$.}
\label{BetaGBMLATComparison_Combined}
\end{figure*}

The results of our joint GBM and LAT spectral fits show that both
softer high-energy power-law spectra and spectral breaks
likely account for the lower-than-expected number of LAT-detected
GRBs.  For the 24 bursts in our spectroscopic subsample where a spectral break is not statistically justified,
the $\beta$ values from the joint fits are systematically softer than
the values found from fitting the GBM data alone.  This may indicate
that the high-energy spectral index for the Band model may in fact be
softer than that deduced from measurements made by previous missions, such
as BATSE, which had a much narrower energy range compared to the
combined coverage of the GBM and LAT.  The GBM+LAT $\beta$ distribution
shown in Figure~\ref{BetaGBMLATComparison_Combined} appears to exclude
the harder spectra found from fits made with just the lower energy
BATSE or GBM data.  In fact, we find no cases of spectra with $\beta >
-2.0$, which would otherwise result in a divergent energy flux at high
energies.

The detection of softer $\beta$ values also provides support for continuum models with multiple components, which have been used to describe novel spectral features detected by the GBM and LAT.  Recent work on bright GRBs by \citet{Guiriec2011} suggests that although the Band function represents many GRB spectra very well in a limited energy range,  it is sometimes possible to discern, even in this limited energy range, contributions such as thermal components in addition to the presumably non-thermal synchrotron emission represented by the Band function.  The addition of such components to a Band function has the effect of modifying the parameter values, in the case of GRB 100724B raising $E_{\rm pk}$ and softening $\beta$ \citep{Guiriec11}.  Whilst these more complex models are not statistically favored in most GRBs due to low photon statistics, their successful fits to some GRBs indicate that the representation of GRB emission by a Band function may be inadequate and lead to overestimates of fluxes when extrapolated to GeV energies.   Because the Band function was developed to represent GRB spectra rather than to parametrize a physical model, it is difficult to decouple physical components from this empirical function, which probably incompletely describes elements of multiple physical phenomena.  Additionally, the superposition of Band functions does not necessarily produce a Band function, so the presence of spectral evolution means that any extrapolation to higher energies from flux-averaged spectra may not be representative of the emission throughout the entire GRB emission period.

\citet{gcd08} have shown that even when integrating over a single spike in a light curve there is a steepening to a softer power-law rather than an exponential cutoff.  This is due to the high-energy power law arising from the sum of instantaneous spectra with an exponential cutoff whose break energy evolves with time.  Likewise, \citet{HD11} have shown that the effect of averaging a time variable opacity cutoff would be manifested as a steepening in the power-law index of the high-energy spectral slope rather than as a sharp cutoff in the spectrum. Likewise, \citet{Baring06} has shown that skin-depth effects tend to smear out exponential attenuation when the source and target photons originate in the same volume, resulting in a similar effect.  Such considerations could explain the softer $\beta$ values found when fitting both the GBM and LAT data, even in cases where a spectral break was not statistically preferred.  Detailed time resolved spectroscopy of bright GBM detected GRBs should be able to discriminate between such pair opacity effects, intrinsically steeper high-energy spectra, or the more complex continuum models discussed above $\beta$ \citep{Guiriec11}.

\begin{figure*}
\includegraphics[width=7in,angle=0]{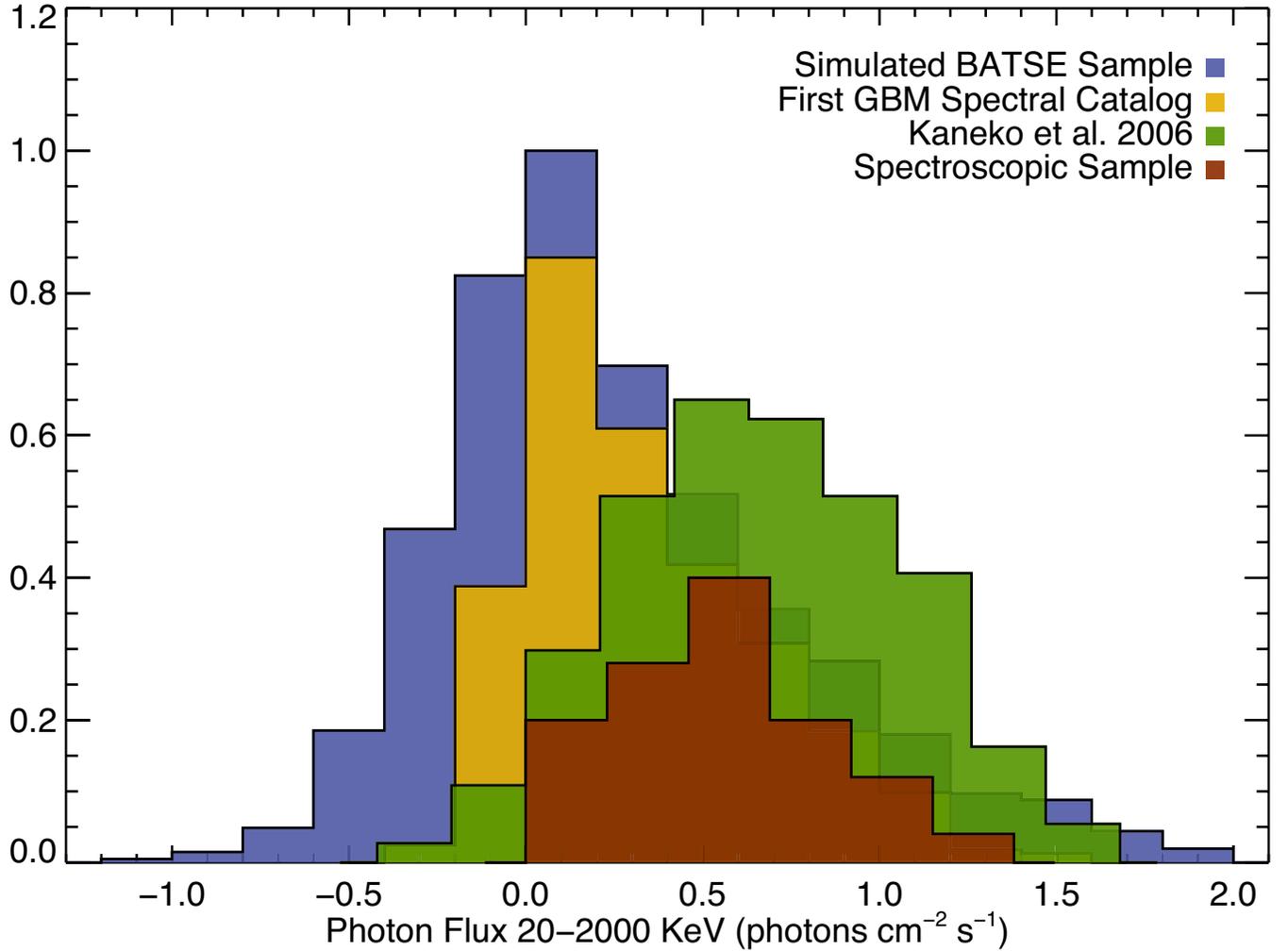}
\caption{The normalized distribution of the time integrated photon flux as determined through our fits to GBM
data for the spectroscopic subsample (red), the bursts in the bright
BATSE catalog presented in \citet{Kaneko06} (green), the bursts that appear in the first GBM spectral catalog (gold), and a sample of
simulated BATSE bursts (blue) using the spectral parameter
distributions given in \citet{Preece00}.  The resulting distributions
show that our spectroscopic subsample is consistent with being drawn
from the distribution of the brightest bursts detected by the GBM and BATSE.}
\label{GoldBrightBGO_SimulatedBaste_FluxDistribution}
\end{figure*}

The bursts in our spectroscopic subsample were chosen specifically because they were among the brightest bursts
detected by the BGO and yet had no appreciable signal in the LAT.
This makes them good candidates to examine for evidence of spectral
breaks, but they may also form a biased data set.  In order to
understand how representative these bursts are of the general GRB population,
we plot in Figure~\ref{GoldBrightBGO_SimulatedBaste_FluxDistribution} the distribution of the time averaged photon flux as determined from fits to GBM
data for bursts in our spectroscopic subsample (red), the bursts which appear in the first GBM spectral catalog (gold) the bursts in the bright
BATSE catalog presented in \citet{Kaneko06} (green), and a sample of
simulated BATSE bursts (blue) using the spectral parameter
distributions given in \citet{Preece00}.  The resulting distributions
show that the spectroscopic subsample is consistent with being drawn
from the distribution of the brightest bursts detected by BATSE.

\begin{figure*}
\includegraphics[width=7in,angle=0]{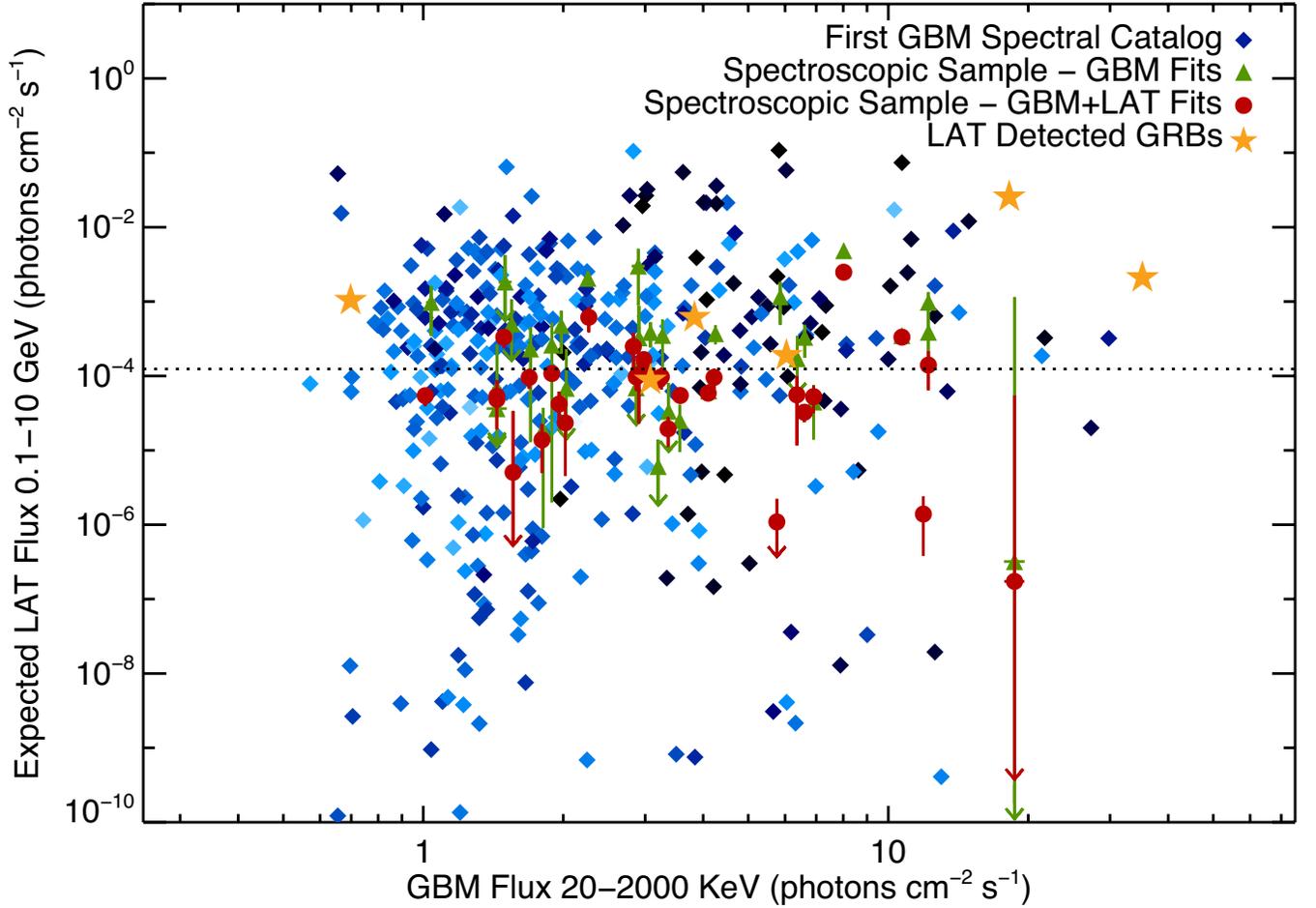}
\caption{Band function model fluxes in the 0.1$-$10\,GeV energy range versus the 0.02$-$2\,MeV energy range for various measure and simulated data.  The gold stars represent the 6 \Fermi\ bursts that were detected by the LAT during the first 18 months that can be well fit by a Band function model; the green circles represent spectral fits to GBM data for the 30 bright BGO bursts in our spectroscopic subsample; the red circles represent spectral fits to GBM and LAT data for the same 30 GRBs; and the blue circles represent bursts that appear in the first GBM spectral catalog for which a Band spectral model could be fit. The color gradient in the GBM sample represents the burst's T90 duration.}
\label{ExpectedFluxVsGBMFluxVsBeta3}
\end{figure*}

We extend this analysis in Figure \ref{ExpectedFluxVsGBMFluxVsBeta3}, where we plot the expected 0.1$-$10\, GeV LAT photon flux versus the 20$-$2000\,keV photon flux for our spectroscopic sample using spectral parameters from the GBM-only fits (green) and from the joint GBM-LAT fits (red), along with the bursts from the first GBM spectral catalog which were in the LAT FOV (blue).  The color gradient in the GBM sample represents the burst's duration, with darker (blue) symbols representing shorter duration bursts.  In addition, we have plotted the 6 LAT-detected bursts (gold) that had spectra that could be fit with a single Band function (i.e., we excluded bursts with extra high-energy components).  The dashed line represents the median T100 upper limit.  The green data points demonstrate how fits to the GBM data without the inclusion of the LAT data yield spectral parameters that over-predict the flux in the LAT energy range, which can be seen by the number of bursts in our spectroscopic subsample that fall above the median upper limit values. The red data points represent the predicted LAT flux for the same GRBs using spectral parameters determined through fits to both the GBM and LAT data. Roughly 50\% of the bursts from the GBM spectral catalog fall above the median T100 upper limit. This would imply that a large fraction of bright GBM detected bursts would have been detectable by the LAT assuming a direct extrapolation of their high-energy spectra.  Therefore, we conclude that intrinsic spectral breaks and/or softer-than-measured high-energy spectra must be fairly common in the GRB population in order to explain the lack of LAT-detected GRBs.

Despite the unknown distances to all but one of the GRBs in our spectroscopic subsample, the allowed range of
$\Gamma_{\gamma\gamma,\max}$ values for $0 < z < 5$ all lie well below$\Gamma_{\gamma\gamma,\max} \sim 720$.  This range of $\Gamma_{\gamma\gamma,\max}$ for the relativistic outflow contrasts with the minimum Lorentz factors that have been calculated for the bright, LAT-detected
GRBs using their highest detected photons.  For GRB~080916C,
GRB~090510, and GRB~09092B, the estimated lower limits for the Lorentz
factors were found to be 887, 1200, and 867 when using single zone models, respectively.  
Therefore, measurements of $\Gamma_{\gamma\gamma,\min}$ and $\Gamma_{\gamma\gamma,\max}$ from both LAT
detections and non-detections reveal a wide distribution in the bulk
Lorentz factor of GRB outflows, with a potential range of over $\sim
10$. 

As discussed above, these estimates of $\Gamma_{\gamma\gamma,\min}$ and $\Gamma_{\gamma\gamma,\max}$ have been calculated using simple single-zone models, which may provide overestimated values compared to time-dependent multi-zone models that take into account the time variability of $\tau_{\gamma\gamma}$. In such a scenario, our estimates of the $\Gamma_{\gamma\gamma,\min}$ and $\Gamma_{\gamma\gamma,\max}$ would need to be rescaled downwards by a factor of 2$-$3, but the large difference between the LAT detected and non-detected GRBs would remain.

\acknowledgements
The \Fermi\ LAT Collaboration acknowledges generous ongoing support
from a number of agencies and institutes that have supported both the
development and the operation of the LAT as well as scientific data analysis.
These include the National Aeronautics and Space Administration and the
Department of Energy in the United States, the Commissariat \`a l'Energie Atomique
and the Centre National de la Recherche Scientifique / Institut National de Physique
Nucl\'eaire et de Physique des Particules in France, the Agenzia Spaziale Italiana
and the Istituto Nazionale di Fisica Nucleare in Italy, the Ministry of Education,
Culture, Sports, Science and Technology (MEXT), high-energy Accelerator Research
Organization (KEK) and Japan Aerospace Exploration Agency (JAXA) in Japan, and
the K.~A.~Wallenberg Foundation, the Swedish Research Council and the
Swedish National Space Board in Sweden.

Additional support for science analysis during the operations phase is gratefully
acknowledged from the Istituto Nazionale di Astrofisica in Italy and the Centre National d'\'Etudes Spatiales in France.

The \Fermi\ GBM collaboration acknowledges support for GBM development, operations and data analysis from NASA in the US and BMWi/DLR in Germany.

\clearpage

\clearpage
\begin{deluxetable}{rrrrrrrrrrrrrrr}
\tablecolumns{11} \tablewidth{0pc}
\tablecaption{Burst Sample with Select Parameters }
\tabletypesize{\small}
\tablehead{ \colhead{GRB} & \colhead{MET\tablenotemark{\dag}} &  \colhead{RA} & \colhead{Dec} &  \colhead{Error} & \colhead{Angle\tablenotemark{\ddag}}   & \colhead{T100} & \colhead{F$_{\rm  lim,T100}$} &  \colhead{F$_{\rm lim,30s}$} & \colhead{F$_{\rm lim, 100s}$}
\\
\colhead{Index} & \colhead{~s~} &  \colhead{~$^{\circ}$~} &  \colhead{~$^{\circ}$~} & \colhead{~$^{\circ}$~}  &\colhead{~$^{\circ}$~} & \colhead{~s~} & \multicolumn{3}{c}{($\times 10^{-5}$ photons cm$^{-2}$ s$^{-1}$)}

}

\startdata
080804972  &     239584816  &      328.70  &      -53.20 &         0.0  &        56.4  &        22.0 &         7.1  &         5.3  &         1.7  \\ 
080805496  &     239630032  &      322.70  &       47.90 &         5.6  &        13.0  &        28.0 &        -  &         2.3  &         0.8  \\
\enddata
\tablenotetext{\dag}{Mission elapsed time relative to January 1, 2001, 0h:0m:0s UTC}
\tablenotetext{\ddag}{Off-axis angle with respect to the LAT boresight}
\label{Table:SampleDefinition} 
\end{deluxetable}

\begin{deluxetable}{cccccc}
\tablecolumns{6} \tablewidth{0pc}
\tabletypesize{\small}
\tablecaption{Spectral Parameters for 30 Bright GBM Detected Bursts - GBM Fits}
\tablehead{ \colhead{GRB} & \colhead{Amplitude} &  \colhead{$\alpha$} & \colhead{$\beta$} & \colhead{$\Epeak$}  & \colhead{C-Stat} 
\\
 \colhead{} & \colhead{($\times 10^{-2}$ photons cm$^{-2}$ s$^{-1}$)} &  \colhead{} & \colhead{} & \colhead{(keV)} & \colhead{}
}
\startdata
080824909  &        0.65  $\pm$        0.33  &       -1.02  $\pm$        0.25  &   -1.84  $\pm$        0.12  &       113.2  $\pm$        47.6  &        1.27  \\ 
080906212  &       12.07  $\pm$        1.58  &       -0.42  $\pm$        0.09  &   -2.38  $\pm$        0.13  &       163.9  $\pm$        11.8  &        1.29  \\ 
080925775  &        1.87  $\pm$        0.19  &       -1.00  $\pm$        0.05  &   -2.13  $\pm$        0.08  &       136.3  $\pm$        11.6  &        1.32  \\ 
081122520  &        4.19  $\pm$        0.44  &       -0.64  $\pm$        0.07  &   -2.44  $\pm$        0.23  &       221.2  $\pm$        19.9  &        1.02  \\ 
081207680  &        0.97  $\pm$        0.04  &       -0.66  $\pm$        0.03  &   -1.98  $\pm$        0.05  &       417.0  $\pm$        24.8  &        2.44  \\ 
081223419  &        4.84  $\pm$        4.20  &       -0.25  $\pm$        0.46  &   -1.85  $\pm$        0.14  &       104.4  $\pm$        33.3  &        1.03  \\ 
081231140  &        1.50  $\pm$        0.08  &       -1.07  $\pm$        0.04  &   -2.59  $\pm$        0.34  &       251.9  $\pm$        20.6  &        1.38  \\ 
090129880  &        0.65  $\pm$        0.10  &       -1.52  $\pm$        0.09  &   -2.31  $\pm$        0.53  &       184.7  $\pm$        62.5  &        1.10  \\
\enddata
\label{Table:SpectralParameters}
\end{deluxetable}

\begin{deluxetable}{ccccc}
\tablecolumns{5} \tablewidth{0pc}
\tabletypesize{\small}
\tablecaption{Measured and Expected Photon Fluxes in the GBM and LAT Bands}
\tablehead{ \colhead{GRB} & \colhead{T90} &  \colhead{Measured Flux 0.02--20 MeV} & \colhead{Expected Flux 0.1--10 GeV}  & \colhead{Flux Limit  0.1--10 GeV} 
\\
 \colhead{} & \colhead{(s)} &  \colhead{(photons cm$^{-2}$ s$^{-1}$)} & \colhead{($\times 10^{-4}$ photons cm$^{-2}$ s$^{-1}$)} & \colhead{($\times 10^{-5}$ photons cm$^{-2}$ s$^{-1}$)}

}

\startdata

080824909  &       28.67  &        1.04  $\pm$        0.04  &        9.75 $\pm$        8.87  &        4.50  \\ 
080906212  &        2.69  &       12.20  $\pm$        0.18  &        3.87 $\pm$        3.32  &       43.60  \\ 
080925775  &       38.14  &        3.08  $\pm$        0.03  &        3.85 $\pm$        2.04  &        5.09  \\ 
081122520  &        4.10  &        6.37  $\pm$        0.12  &        1.71 $\pm$        2.49  &       24.75  \\ 
081207680  &      104.45  &        2.26  $\pm$        0.02  &       20.50 $\pm$        6.49  &        4.31  \\ 
081223419  &        2.36  &        2.90  $\pm$        0.13  &       30.00 $\pm$       29.70  &       34.95  \\ 
081231140  &       27.65  &        3.37  $\pm$        0.04  &        0.34 $\pm$        0.66  &        2.49  \\ 
090129880  &       16.38  &        2.03  $\pm$        0.05  &        0.68 $\pm$        2.26  &        6.94  \\
\enddata
\label{Table:ExpectedFlux}
\end{deluxetable}

\begin{deluxetable}{ccccccccc}
\tablecolumns{9} \tablewidth{0pc}
\tabletypesize{\small}
\rotate 
\tablecaption{Spectral Parameters for 30 Bright GBM Detected Bursts - GBM $\&$ LAT Fits}
\tablehead{ \colhead{GRB} & \colhead{Amplitude} &  \colhead{$\alpha$} & \colhead{$\beta$} & \colhead{$\Epeak$}  & \colhead{Cash} & \colhead{DOF} & \colhead{$\Delta\beta$}  & \colhead{$\Delta$C-Stat} 
\\
 \colhead{} & \colhead{($\times 10^{-2}$ photons cm$^{-2}$ s$^{-1}$)} &  \colhead{} & \colhead{} & \colhead{(keV)} & \colhead{} & \colhead{} & \colhead{} & \colhead{}

}

\startdata

080824909  &        0.53  $\pm$        0.12  &       -1.11  $\pm$        0.13  &      -2.41  $\pm$        0.15  &       151.2  $\pm$        31.3  &      475.73  &        378  &       -0.56  $\pm$        0.15  &        2.32  \\ 
080906212  &       10.79  $\pm$        1.10  &       -0.48  $\pm$        0.07  &      -2.55  $\pm$        0.12  &       176.4  $\pm$        10.1  &      639.11  &        504  &       -0.17  $\pm$        0.12  &        0.20  \\ 
080925775  &        1.70  $\pm$        0.12  &       -1.05  $\pm$        0.04  &      -2.42  $\pm$        0.09  &       154.0  $\pm$         9.4  &      500.77  &        380  &       -0.30  $\pm$        0.09  &        9.53  \\ 
081122520  &        3.99  $\pm$        0.35  &       -0.67  $\pm$        0.07  &      -2.64  $\pm$        0.19  &       232.7  $\pm$        17.0  &      501.82  &        502  &       -0.20  $\pm$        0.19  &        0.04  \\ 
081207680  &        0.87  $\pm$        0.02  &       -0.73  $\pm$        0.02  &      -2.70  $\pm$        0.12  &       528.2  $\pm$        24.3  &      952.43  &        385  &       -0.72  $\pm$        0.12  &       14.57  \\ 
081223419  &        2.57  $\pm$        1.02  &       -0.57  $\pm$        0.25  &      -2.23  $\pm$        0.15  &       149.6  $\pm$        32.0  &      399.95  &        380  &       -0.37  $\pm$        0.15  &        0.67  \\ 
081231140  &        1.46  $\pm$        0.07  &       -1.08  $\pm$        0.03  &      -3.46  $\pm$        0.71  &       265.7  $\pm$        16.0  &      509.18  &        378  &       -0.88  $\pm$        0.71  &        0.01  \\ 
090129880  &        0.61  $\pm$        0.07  &       -1.55  $\pm$        0.07  &      -4.16  $\pm$       36.00  &       219.9  $\pm$        56.2  &      406.74  &        379  &       -1.85  $\pm$       36.00  &        0.00  \\
\enddata
\label{Table:GBMLAT_SpectralFitParameters}
\end{deluxetable}

\bibliography{ms.bib}

\end{document}